# Revealing Hidden Orbital Pathways in Non-Thermal Hot Carrier Relaxation of MXenes via Non-Secular Redfield Quantum Kinetics

Ali Asghar Molavi Choobini[1*] Abbas Chimeh[1,2], Jinhui Zhong[3]
[1]Quantum Matter Lab, Department of Physics, College of Science, University of Tehran, Tehran 14399-55961, Iran,
[2]Nexus for Quantum Coherence and Entanglement in Light-Matter Systems (Qcelms), University of Tehran, P.O. Box 14395-547, Tehran, Iran,
[3]Department of Materials Science and Engineering, Southern University of Science and Technology, Shenzhen 518055, China

**Abstract:** Non-thermal carrier relaxation is routinely inferred from population dynamics or spectroscopic observables, yet neither class of quantity uniquely identifies the microscopic channels through which energy and coherence are redistributed. We introduce a pathway-resolved quantum-kinetic framework that simultaneously projects ultrafast relaxation onto orbital populations, directional inter-orbital transferes, coherence, spectroscopic visibility, and a hidden-pathway sector of the dynamical transfer network. Application to MXenes exposes strongly non-uniform orbital redistribution together with material-specific hierarchies of microscopic transfer channels. Temperature, excitation amplitude, and dissipative parameters modulate pathway competition and spectral amplitudes while leaving the identity of the dominant channels largely intact. Instantaneous transfer contributions, cumulative directional transfer, coherence, and spectroscopic visibility are shown to follow inequivalent hierarchical orderings. This non-equivalence isolates a set of hidden pathways that remain dynamically consequential despite weak conventional spectroscopic signatures. The resulting time–energy–coherence representation recasts non-thermal relaxation as a structured dynamical network comprising observable and hidden sectors, thereby providing a general methodology for resolving microscopic orbital-transfer pathways in driven quantum materials.

*Corresponding author: E-mail address: aa.molavich@ut.ac.ir

## I. Introduction

Hot-carrier dynamics in two-dimensional (2D) MXenes have emerged as an important topic in condensed-matter physics and nanophotonics, driven by their combination of metallic conductivity, broadband plasmonic response extending from the visible to the near-infrared, and chemically tunable surface terminations (–O, –OH, –F) [1–3]. In prototypical systems such as $Ti_3C_2T_x$, plasmon decay generates strongly nonequilibrium carriers whose subsequent evolution involves competing femtosecond-scale processes, including electron–electron scattering, non-thermalized electron transfer (NET), non-thermalized electron-induced heat transfer (NEIHT), and electron–phonon coupling [4–7]. Real-time spectroscopic studies of MXene/molecule interfaces have further revealed a competition between coherent interfacial charge transfer and incoherent thermalization, with the resulting dynamics governed by the electronic density of states near the Fermi level, orbital hybridization, and non-adiabatic coupling. These observations position MXenes as a useful platform for probing nonequilibrium carrier dynamics and for testing quantum-kinetic descriptions that explicitly resolve orbital degrees of freedom in ultrafast light–matter interactions. Although we focus on MXenes as representative metallic 2D systems with rich *d–p* hybridization, the framework developed below is constructed to be transferable to other orbital-active 2D quantum materials. The technological relevance of these nonequilibrium dynamics spans photocatalysis, ultrafast photonics, and energy conversion. Non-thermalized hot electrons in MXenes can participate directly in redox processes, while MXene-based saturable absorbers and all-optical modulators provide sub-200-fs optical responses. Plasmon-enhanced hot-carrier injection into semiconductor heterostructures further enables broadband and wavelength-tunable photodetection. Other applications exploit the efficient conversion of optical energy into heat for solar desalination, photothermal therapy, and thermal management in microelectronic systems [8–10]. Hot-carrier-assisted charge transfer has likewise been explored for accelerating reaction kinetics in lithium–sulfur batteries, supercapacitors, and electrocatalytic energy-storage systems. These diverse applications motivate a quantitative understanding of how nonequilibrium carrier energy is redistributed among electronic, vibrational, and interfacial degrees of freedom.

Recent advances in plasmonic nanomaterials and two-dimensional systems have established interface engineering, through surface-termination control, heterostructure design, and compositional tuning, as an effective strategy for modulating hot-carrier lifetimes, transport pathways, and energy distributions. Reviews of MXene plasmonics highlight the central role of localized surface plasmon resonances in solar photothermal conversion, hydrogen production, desalination, and environmental remediation [11], while the optical transparency, metallic conductivity, and tunable optoelectronic response of $Ti_3C_2T_x$ have been emphasized for photonic and electronic applications [12]. Integration with conventional semiconductors also improves charge transport by providing transparent electrodes and functional interfacial layers [13]. Ultrafast charge and thermal transport under optical excitation are controlled by surface terminations and electronic structure [14], although LSPR-induced hot-hole dynamics remain less explored than hot-electron dynamics, despite their relevance to oxidative photocatalysis [15]. Microscopic studies have shed light on the multichannel nature of charge transport and hot-carrier relaxation in MXenes. Transport involves high-mobility intra-flake conduction and field-dependent inter-flake hopping with Poole–Frenkel characteristics [16], while synthesis conditions strongly influence structural quality and ultrafast carrier dynamics [17]. Surface termination affects electron–phonon coupling in MBenes, where non-terminated MBenes show slower carrier relaxation [18]. The shrinking of the lateral dimensions of $Ti_3C_2T_x$ enhances electron–phonon coupling and expedites thermal diffusion, thereby favouring hot-carrier cooling [19]. Nonthermal pathways, including direct plasmon-to-coherent-phonon coupling that bypasses electron–electron thermalization, have also been identified [20]. Photothermal effects can dominate charge transport and produce persistent negative photoconductivity when heat dissipation is slow [21]. In $Ti_3C_2T_x$, carrier cooling proceeds on picosecond timescales without an apparent hot-phonon bottleneck, whereas lattice heat dissipation extends over hundreds of nanoseconds [22]. Collectively, these studies indicate that hot-carrier relaxation in MXenes is an intrinsically multichannel and state-dependent process. Interfacial coupling adds further complexity by introducing electronic pathways that can compete with intrinsic carrier relaxation. Hot-electron injection across $Ti_3C_2T_x$ Schottky interfaces occurs on tens-of-femtoseconds timescales [23], while charge transfer and secondary excitation in $Ti_3C_2T_x$–$MoS_2$ heterostructures proceed within sub-150 fs [24]. S-scheme/Schottky heterojunctions enhance charge separation and catalytic activity through synergistic interfacial coupling [25], and efficient charge transfer in MXene/g-$C_3N_4$ systems correlates with increased photosensitivity and prolonged carrier lifetimes [26]. Theoretical frameworks now incorporate both thermal and nonthermal contributions to plasmonic hot-carrier dynamics [27]. Real-time TDDFT calculations show that plasmon-generated hot electrons in $Ti_3C_2T_x$ substantially lower reaction barriers, including those for $H_2O$ dissociation [28], while atomistic simulations reveal preferential hot-carrier localization at low-coordinated surface sites [29]. Charge-transfer resonances further determine sensitivity and selectivity in MXene-based surface-enhanced Raman spectroscopy [30].

Despite substantial advances in the characterization of hot-carrier dynamics, prevailing approaches typically describe relaxation through energy-resolved populations, scattering rates, effective temperatures, or spectroscopic observables. Although these quantities provide complementary information on carrier cooling, they typically constitute inequivalent projections of the underlying quantum dynamics and do not uniquely identify the microscopic channels responsible for energy redistribution. Orbital populations register the net balance of incoming and outgoing transfer, directional transfer quantifies the instantaneous dynamical strength of individual channels, coherence encodes interference among quantum pathways, and spectroscopic cross-peaks furnish a visibility-weighted representation of the evolving dynamics. Distinct microscopic pathways may therefore generate closely similar energy-resolved population traces while differing markedly in orbital composition, inter-orbital coherence, branching topology, and spectroscopic visibility; a dynamically consequential pathway need not appear spectroscopically prominent. Here, this incompleteness is addressed through a pathway-resolved quantum-kinetic framework that treats orbital character as a dynamical degree of freedom rather than a static label and resolves non-equilibrium relaxation simultaneously across coupled energy, momentum, orbital, and coherence spaces. The framework maps relaxation onto a network of coherent and dissipative inter-orbital transitions and reconstructs its evolution through complementary representations of orbital populations, inter-orbital energy flow, coherence-resolved spectroscopic proxy, and a hidden-pathway sector defined by the dynamical topology of the transfer network. It distinguishes pathway strength from pathway visibility and classifies microscopic channels according to temporal accumulation, energetic localization, coherence content, network connectivity, and spectroscopic observability. The resulting analysis isolates hidden orbital relaxation pathways that population dynamics or spectroscopic intensity alone cannot identify and establishes a microscopic connection between orbital structure, energy redistribution, and experimentally measurable relaxation signatures. Applied to MXenes, the framework resolves how temperature, excitation conditions, and dissipative quantum dynamics reshape the hierarchy of microscopic relaxation channels. More broadly, it provides a systematic route for identifying material-specific pathways of nonequilibrium energy redistribution and for examining how orbital structure, surface chemistry, and interfacial coupling influence their connectivity, coherence, and observability, thereby providing a basis for pathway-selective control of non-thermal dynamics in two-dimensional quantum materials.

# II. Theoretical Framework

## 2.1 Correlated Orbital Hamiltonian

The coherent electronic structure is formulated in a localized transition-metal $d$-orbital basis, in which the orbital connectivity is resolved explicitly. The total Hamiltonian is written as:

$$H(t) = H_{\mathrm{el}} + H_{\mathrm{Pump}}(t), \quad (1)$$

where the time-independent correlated electronic Hamiltonian reads

$$H_{\mathrm{el}} = H_{\mathrm{CF}} + H_{\mathrm{hyb}} + H_{\mathrm{hop}} + H_{\mathrm{int}} + H_{\mathrm{SOC}}. \quad (2)$$

The successive terms describe the local crystal-field landscape, intra-site orbital hybridization, inter-site hopping, local Coulomb correlations, and spin-orbit coupling, respectively. Together they define the correlated electronic manifold in which the subsequent nonequilibrium quantum dynamics evolve. The light-matter interaction that generates the initial non-thermal hot-carrier distribution is described within the electric-dipole approximation by:

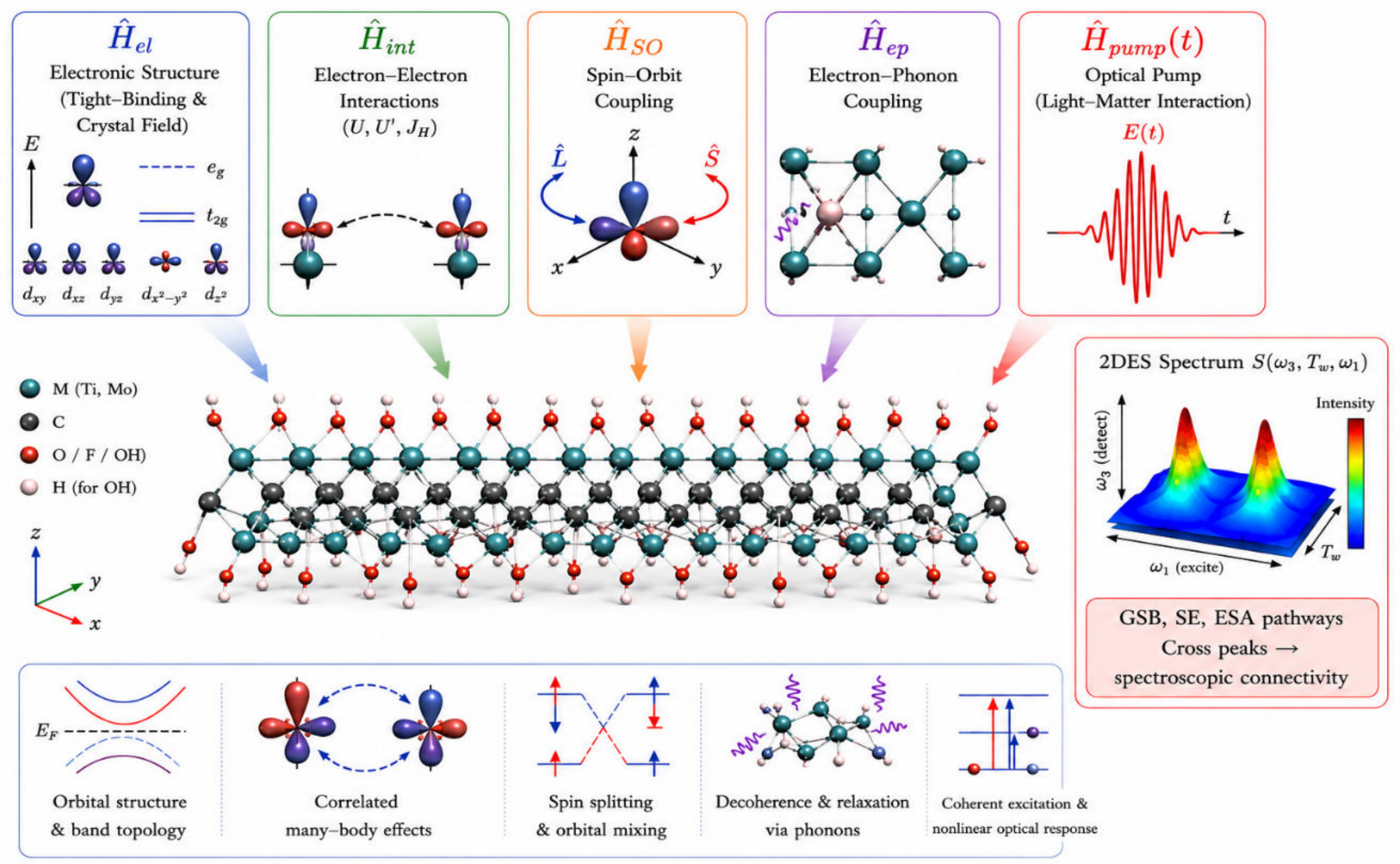

**Figure 1:** Microscopic Hamiltonian framework for MXenes. Schematic of the Hamiltonian comprising electronic structure ($\widehat{H}_{el}$), electron–electron interactions ($\widehat{H}_{int}$), spin–orbit coupling ($\widehat{H}_{SOC}$), electron–phonon coupling ($\widehat{H}_{ep}$), and optical driving ($\widehat{H}_{pump}(t)$).

$$H_{\text{Pump}}(t) = -\boldsymbol{\mu} \cdot \mathbf{E}(t), \tag{3}$$

where $\mathbf{E}(t)$ is the electric field of the ultrashort optical pulse and $\boldsymbol{\mu}$ is the electric-dipole operator expressed in the localized transition-metal $d$-orbital basis. The matrix elements of $\boldsymbol{\mu}$ encode the optically allowed orbital transitions that launch the subsequent non-equilibrium dynamics. After the pump pulse has passed, $H_{\text{Pump}}(t)$ vanishes and the system evolves under the remaining time-independent Hamiltonian according to the non-secular Redfield quantum-kinetic equation. The local crystal-field contribution is given by

$$H_{\text{CF}} = \sum_{Rm\sigma} \epsilon_m^{(R)} c_{Rm\sigma}^{\dagger} c_{Rm\sigma}, \tag{4}$$

where $R$ labels the lattice site, $m$ denotes a localized orbital, and $\sigma$ is the spin projection. The onsite energy $\epsilon_m^{(R)}$ incorporates the local electrostatic environment and crystal-field potential experienced by orbital $m$. The corresponding orbital splitting is $\Delta_{\text{CF}}^{mn} = \epsilon_m - \epsilon_n$, which sets the local energetic separation between orbital sectors and thus determines their relative accessibility during nonequilibrium carrier redistribution. Surface termination, local structural distortions, and chemical composition can therefore modify the accessible orbital landscape through changes in $\epsilon_m$. Intra-site orbital mixing is described by:

$$H_{\text{hyb}} = \sum_R \sum_{m \neq n} \sum_\sigma V_{mn}^{(R)} c_{Rm\sigma}^{\dagger} c_{Rn\sigma}, \tag{5}$$

where $V_{mn}^{(R)}$ is the hybridization matrix element between orbitals $m$ and $n$ on the same site. Unlike $H_{\text{CF}}$, which determines the local orbital energies, $H_{\text{hyb}}$ generates coherent mixing between orbital sectors. A finite $V_{mn}^{(R)}$ therefore allows an electronic eigenstate to acquire weight from multiple localized orbitals, providing the local coherent connectivity through which intermediate orbital character can emerge during nonequilibrium evolution. Spatial propagation is introduced through the inter-site hopping Hamiltonian:

$$H_{\mathrm{hop}} = \sum_R \sum_\ell \sum_{\boldsymbol{\delta}_\ell} \sum_{mn\sigma} \left[t_{mn}^{(\ell)} c_{Rm\sigma}^\dagger c_{R+\boldsymbol{\delta}_\ell,n\sigma} + \mathrm{H.c.}\right], \tag{6}$$

where $\boldsymbol{\delta}_\ell$ denotes the bond vectors connecting sites within the $\ell$th coordination shell and $t_{mn}^{(\ell)}$ is the corresponding inter-site hopping amplitude between orbitals $m$ and $n$. The explicit Hermitian-conjugate term ensures Hermiticity without imposing a separate bond-orientation convention. In a translationally invariant representation, the hopping contribution generates the momentum-dependent orbital matrix $h_{mn}(\mathbf{k}) = \sum_\ell \sum_{\boldsymbol{\delta}_\ell} t_{mn}^{(\ell)} e^{i\mathbf{k}\cdot\boldsymbol{\delta}_\ell}$, which connects the localized orbital basis to the electronic band dispersion. Thus, $H_{\mathrm{hyb}}$ determines local orbital mixing, whereas $H_{\mathrm{hop}}$ extends this connectivity across the lattice and determines how orbital character is distributed throughout momentum space. Local correlations are incorporated using the rotationally invariant Kanamori interaction,

$$\begin{aligned} H_{\mathrm{int}} = \; & U \sum_{Rm} n_{Rm\uparrow} n_{Rm\downarrow} + U' \sum_R \sum_{m<n} n_{Rm} n_{Rn} \\ & -J_{\mathrm{H}} \sum_R \sum_{m<n} \left(2\mathbf{S}_{Rm} \cdot \mathbf{S}_{Rn} + \tfrac{1}{2} n_{Rm} n_{Rn}\right) + J_{\mathrm{H}} \sum_R \sum_{m\neq n} c_{Rm\uparrow}^\dagger c_{Rm\downarrow}^\dagger c_{Rn\downarrow} c_{Rn\uparrow}, \end{aligned} \tag{7}$$

where $n_{Rm\sigma} = c_{Rm\sigma}^\dagger c_{Rm\sigma}$, $n_{Rm} = \sum_\sigma n_{Rm\sigma}$ and the orbital-resolved spin operator is $\mathbf{S}_{Rm} = \frac{1}{2}\sum_{\sigma\sigma'} c_{Rm\sigma}^\dagger \boldsymbol{\sigma}_{\sigma\sigma'} c_{Rm\sigma'}$, with $\boldsymbol{\sigma}$ denoting the vector of Pauli matrices. Here, $U$ is the intra-orbital Coulomb repulsion, $U'$ is the inter-orbital density interaction, and $J_{\mathrm{H}}$ is Hund's exchange coupling. The first two terms penalize double and inter-orbital occupancy, respectively, whereas the exchange and pair-hopping terms account for Hund-coupled spin configurations and correlated transfer of electron pairs between orbitals. For a rotationally invariant interaction within a degenerate $d$-orbital manifold, the Kanamori parameters satisfy $U' = U - 2J_{\mathrm{H}}$. When the effective interaction parameters are obtained independently from a material-specific downfolding procedure, $U'$ and $J_{\mathrm{H}}$ may be retained as independent effective parameters, in which case the rotationally invariant relation is not imposed. Within this correlated orbital Hamiltonian the individual terms play complementary roles: $H_{\mathrm{CF}}$ determines the local orbital-energy landscape, $H_{\mathrm{hyb}}$ introduces coherent mixing between orbitals on the same site, $H_{\mathrm{hop}}$ extends this connectivity across the lattice and into momentum space, and $H_{\mathrm{int}}$ couples distinct electron configurations through Coulomb and Hund interactions. These contributions therefore should not be interpreted as independent relaxation mechanisms; rather, together they define the correlated electronic manifold within which nonequilibrium dynamics take place. The hybridization and hopping matrix elements $V_{mn}^{(R)}$ and $t_{mn}^{(\ell)}$ establish coherent pathways through which an initially excited electronic state can acquire amplitude in intermediate orbital sectors, while the interaction parameters $U$, $U'$, and $J_{\mathrm{H}}$ modify the energetic and character of the configurations participating in this redistribution. Irreversible population relaxation and decoherence arise only when this correlated electronic manifold is coupled to dissipative degrees of freedom, most notably lattice vibrations through $H_{\mathrm{e-ph}}$. The resulting nonequilibrium evolution can therefore redistribute carrier population and energy through sequences of orbital sectors that are not reducible to a single direct transition between the initially and finally populated states. This distinction between coherent electronic connectivity and irreversible environmental relaxation is central to the identification of hidden orbital pathways: the former determines which microscopic configurations can communicate, whereas the latter determines how population, energy, and coherence are irreversibly redistributed among them. Spin–orbit coupling is incorporated into the correlated electronic Hamiltonian as:

$$H_{\mathrm{SOC}} = \lambda_{SOC}\, \mathbf{L} \cdot \mathbf{S}, \qquad \mathbf{S} = \frac{\hbar}{2}\boldsymbol{\sigma}, \tag{8}$$

where $\lambda_{SOC}$ is the spin–orbit coupling strength, $\mathbf{L}$ is the orbital angular-momentum operator, and $\boldsymbol{\sigma} = (\sigma_x, \sigma_y, \sigma_z)$ denotes the vector of Pauli matrices. By coupling the orbital and spin degrees of freedom, $H_{\mathrm{SOC}}$ modifies the composition and energies of the correlated electronic eigenstates. Consequently, it alters the transition matrix elements and the coherence dynamics that enter the subsequent nonequilibrium evolution governed by the full electronic Hamiltonian, $H_{\mathrm{el}}|\beta\rangle = E_\beta|\beta\rangle$, where $|\beta\rangle$ denotes a correlated electron spin–orbital eigenstate and $E_\beta$ its corresponding eigenenergy. The index $\beta$ refers to the complete electronic eigenstate and is therefore distinct from the localized-orbital indices used in the Hamiltonian and from the phonon indices introduced below. The one-electron matrix elements entering $H_{\mathrm{CF}}$, $H_{\mathrm{hyb}}$, and $H_{\mathrm{hop}}$ are obtained from a first-principles Wannier representation, while the interaction parameters and the SOC matrix are determined independently from the corresponding first-principles parameterization and projected onto the same localized basis. This procedure preserves a consistent representation of the orbital, spatial, spin, and correlation structure of each material; the material-specific numerical values and their computational provenance are provided in the Supplementary Information. The eigenstate spectrum defines the electronic transition energy and Bohr frequency between states $|\beta\rangle$ and $|\alpha\rangle$ as $\Delta E_{\alpha\beta} = E_\alpha - E_\beta = \hbar\omega_{\alpha\beta}$. For $\Delta E_{\alpha\beta} > 0$, $\omega_{\alpha\beta}$ corresponds to the upward transition $\beta \to \alpha$, while the reverse process is described by $\omega_{\beta\alpha} = -\omega_{\alpha\beta}$. In the absence of dissipative coupling, an electronic coherence evolves according to $\rho_\beta(t) = \rho_{\alpha\beta}(0)\exp(-i\omega_{\alpha\beta}t)$, so that the Bohr frequencies set both the intrinsic phase-evolution scales of the electronic coherences and the energy-matching conditions entering phonon-assisted transitions. Because the eigenstates $|\alpha\rangle$ are generally coherent superpositions of localized orbital and spin configurations, an eigenstate-to-eigenstate transition cannot, in general, be identified with a unique localized orbital transition. Its microscopic character is distributed among the orbital components of the participating eigenstates. As a result, a population transfer that appears as a single transition in an energy-resolved description may contain multiple orbital contributions, including pathways mediated by intermediate orbital sectors. Retaining the orbital composition of the correlated eigenstates provides the basis for resolving orbital-selective relaxation pathways. Irreversible electronic relaxation is introduced by coupling the correlated electronic subsystem to lattice vibrations. The phonon bath is described as a set of harmonic modes:

$$H_{\mathrm{ph}} = \sum_{\mathbf{q}\eta} \hbar\omega_{\mathbf{q}\eta} \left(b_{\mathbf{q}\eta}^\dagger b_{\mathbf{q}\eta} + \frac{1}{2}\right), \quad (9)$$

where $\mathbf{q}$ and $\eta$ denote the phonon wavevector and branch, respectively, and $\omega_{\mathbf{q}\eta}$ is the corresponding phonon frequency. The electron–phonon interaction is written in the localized orbital basis as:

$$H_{\mathrm{e-ph}} = \frac{1}{\sqrt{N_R}} \sum_{\mathbf{k},\mathbf{q},mn\sigma,\eta} g_{mn}^\eta(\mathbf{k},\mathbf{q})\, c_{\mathbf{k}+\mathbf{q},m\sigma}^\dagger c_{\mathbf{k},n\sigma} \left(b_{\mathbf{q}\eta} + b_{-\mathbf{q}\eta}^\dagger\right), \quad (10)$$

where $g_{mn}^\eta(\mathbf{k},\mathbf{q})$ denotes the electron-phonon matrix element for the corresponding orbital and phonon states. Upon transformation to the correlated electronic eigenbasis, the coupling is characterized by matrix elements $G_{\mu\nu}^\eta(\mathbf{k},\mathbf{q})$, which retain the orbital composition of the initial and final correlated states. When obtained from first-principles lattice perturbations, these matrix elements are evaluated as $G_{\alpha\beta}^\eta(\mathbf{k},\mathbf{q}) = \langle \alpha, \mathbf{k}+\mathbf{q} | \delta_{\mathbf{q}\eta} V_{\mathrm{KS}} | \beta, \mathbf{k} \rangle$, where $\delta_{\mathbf{q}\eta} V_{\mathrm{KS}}$ is the first-order phonon-induced variation of the self-consistent potential. The corresponding transition-resolved electron–phonon spectral density is defined as:

$$J_{\alpha\beta}(\omega) = \frac{1}{N_R} \sum_{\mathbf{k},\mathbf{q},\eta} \left|G_{\alpha\beta}^\eta(\mathbf{k},\mathbf{q})\right|^2 \delta(\omega - \omega_{\mathbf{q}\eta}). \quad (11)$$

Thus, $J_{\alpha\beta}(\omega)$ quantifies the phonon spectral weight coupled to the $\beta \rightarrow \alpha$ electronic transition. Its overlap with the electronic transition frequencies, together with the corresponding thermal occupation factors, determines the dissipative phase space entering the Redfield relaxation kernel. The resulting relaxation channels are governed microscopically by the electronic energy separations, orbital composition, electron–phonon matrix elements, phonon spectrum, and thermal occupation rather than by phenomenological relaxation times. A soft temporal cutoff is introduced only as a numerical regularization of the finite-time evaluation of the bath correlation transform. It does not represent a physical finite-memory kernel and does not modify the time-local Born–Markov structure of the Redfield equation. The resulting dissipative dynamics therefore remain Markovian, with the cutoff controlling numerical convergence rather than introducing an additional physical memory timescale.

## 2.2 Redfield Quantum kinetics of Nonequilibrium Relaxation

The nonequilibrium electronic dynamics are described within a reduced density-matrix framework in which the correlated spin–orbital electronic manifold is coupled to a phonon environment. The total Hamiltonian is written as:

$$H_{\mathrm{tot}}(t) = H_{\mathrm{el}} + H_{\mathrm{ph}} + H_{\mathrm{e-ph}} + H_{\mathrm{drive}}(t), \quad (12)$$

where $H_{\mathrm{el}}$ is the correlated electronic Hamiltonian defined above, $H_{\mathrm{ph}}$ describes the harmonic phonon bath, $H_{\mathrm{e-ph}}$ mediates electron–phonon scattering, and $H_{\mathrm{drive}}(t)$ describes the external optical excitation. This separation distinguishes the optical preparation of the nonequilibrium state from its subsequent dissipative evolution. The density operator of the combined electron–phonon system obeys the Liouville–von Neumann equation, $\dot{\rho}_{\mathrm{tot}}(t) = -i[H_{\mathrm{tot}}(t), \rho_{\mathrm{tot}}(t)]/\hbar$, while the electronic state is described by the reduced density matrix $\rho(t) = \mathrm{Tr}_{\mathrm{ph}}[\rho_{\mathrm{tot}}(t)]$. In the eigenbasis $\{|\mu\rangle\}$ of $H_{\mathrm{el}}$, its diagonal elements describe electronic populations and its off-diagonal elements describe quantum coherences. Retaining both is essential because the orbital redistribution analyzed here can originate from coherent mixing as well as irreversible population transfer. For weak electron–phonon coupling, the interaction is treated to second order, while the phonon bath is assumed to remain close to thermal equilibrium. When the bath correlation time $\tau_{\mathrm{B}}$ is short compared with the characteristic electronic relaxation time $\tau_{\mathrm{R}}$, i.e., $\tau_{\mathrm{B}} \ll \tau_{\mathrm{R}}$, the Born–Markov approximation yields a time-local kinetic equation. In the interaction representation of the electron–phonon coupling, the bath enters through equilibrium two-point correlation functions $C_{ab}(\tau) = \langle B_a(\tau)B_b(0)\rangle_{\mathrm{ph}}$ for a coupling of the form $H_{\mathrm{e-ph}} = \sum_a A_a \otimes B_a$. Their one-sided Fourier transforms define the complex Redfield kernels $\Gamma_{ab}(\omega) = \hbar^{-2}\int_0^\infty d\tau\, e^{i\omega\tau} C_{ab}(\tau)$. The real parts of these kernels determine the dissipative transition contributions, whereas the imaginary parts give bath-induced energy renormalization. In the present treatment, the corresponding energy renormalization is retained consistently with the chosen Redfield convention. The reduced dynamics are therefore propagated in the non-secular Redfield form:

$$\frac{d\rho_{\alpha\beta}(t)}{dt} = -i\omega_{\alpha\beta}\rho_{\alpha\beta}(t) - \sum_{\kappa\lambda} R_{\alpha\beta,\kappa\lambda}\rho_{\kappa\lambda}(t), \quad (13)$$

where $R_{\alpha\beta,\kappa\lambda}$ is the Redfield relaxation tensor constructed from the electron–phonon matrix elements and the bath correlation functions. In the adopted index convention, the tensor may be written as:

$$R_{mn,pq} = \Gamma^{(+)}_{qn,mp} + \Gamma^{(-)}_{qn,mp} - \delta_{np}\sum_r \Gamma^{(+)}_{mr,rq} - \delta_{mq}\sum_r \Gamma^{(-)}_{nr,rp}, \quad (14)$$

with the one-sided transition kernels that are determined by the bath correlation functions:

$$\Gamma^{(+)}_{mn,pq} = \frac{1}{\hbar^2}\int_0^\infty C_{mn,pq}(\tau)e^{-i\omega_{pq}\tau}\, d\tau \quad (15a)$$

$$\Gamma^{(-)}_{mn,pq} = \frac{1}{\hbar^2}\int_0^{\infty} C^*_{mn,pq}(\tau)e^{i\omega_{mn}\tau}\, d\tau \tag{15b}$$

These expressions make explicit that the Redfield tensor is determined by microscopic bath correlations rather than by phenomenological relaxation or dephasing times. The baseline relaxation dynamics are determined by the correlated electronic Hamiltonian, electron–phonon matrix elements, bath spectral density, and temperature entering the Redfield tensor. Additional parameters introduced in the parameter-sweep analysis are used only for controlled sensitivity tests of the resulting dynamics and are not part of the baseline microscopic parameterization. In particular, the reference relaxation scale, pure-dephasing coefficient, non-secular weighting, and secular-mixing parameter are varied independently around the baseline calculation to assess the robustness of the orbital-resolved dynamics. The Redfield tensor is constructed once in the eigenbasis of the field-free correlated electronic Hamiltonian and is kept fixed during the optical excitation. The time dependence of the Liouvillian therefore arises solely from the explicitly time-dependent dipole interaction, while the dissipative kernel is not recomputed in an instantaneous field-dressed basis. The dissipative contribution acts throughout the propagation, including during the pump pulse. The imaginary parts of the Redfield kernels are retained as the corresponding bath-induced energy-renormalization (Lamb-shift) contribution, consistent with Eqs. (15a)–(15b). The non-secular form is retained because the objective is to resolve orbital pathways that may involve transient coherences. When different electronic transitions have comparable Bohr frequencies, dissipative coupling between different density-matrix elements can remain relevant on the relaxation timescale. A secular approximation would remove such terms and could therefore suppress coherence–population and coherence–coherence couplings that contribute to transient orbital redistribution. The use of the non-secular Redfield equation is consequently restricted to the weak-coupling and short-memory regime specified above. During numerical propagation, the trace and Hermiticity of the reduced density matrix are monitored, together with its instantaneous eigenvalues. No positivity correction, eigenvalue clipping, or projection onto the positive semidefinite cone is applied during the propagation. The minimum eigenvalue is recorded throughout each simulation to assess any loss of positivity associated with the non-secular Redfield approximation. For the electron–phonon interaction considered here, the system operators entering the Redfield tensor are determined by the phonon-induced electronic matrix elements $G^{\eta}_{\alpha\kappa}(\mathbf{k},\mathbf{q})$ introduced above. These matrix elements retain the orbital composition of the correlated spin–orbital eigenstates and therefore determine the orbital character of each phonon-assisted transition. The corresponding dissipative contributions are weighted by the phonon spectral density and by the thermal occupation of the bath. In particular, phonon emission and absorption are weighted by $n_{\mathrm{B}}(\omega) + 1$ and $n_{\mathrm{B}}(\omega)$, respectively, where $n_{\mathrm{B}}(\omega) = [\exp(\hbar\omega/k_{\mathrm{B}}T_{\mathrm{ph}}) - 1]^{-1}$. The efficiency of a microscopic relaxation channel is governed jointly by the electronic energy separation, electron–phonon matrix element, phonon spectral weight, and available thermal phase space, rather than by a single phenomenological relaxation time. The Redfield evolution provides the full time-dependent electronic density matrix from which the orbital-resolved observables are constructed. For a correlated electronic manifold, the orbital population is defined through the occupation operator, $\hat{n}_i = \sum_{\sigma} c^{\dagger}_{i\sigma}c_{i\sigma}$, where $i$ denotes the localized orbital sector and the sum runs over the corresponding spin states. The instantaneous orbital population is then:

$$P_i^{orb}(t) = Tr_{el}[\rho(t)\hat{n}_i] = \sum_{\alpha\beta}\rho_{\alpha\beta}(t)\langle\beta|\hat{n}_i|\alpha\rangle \tag{16}$$

This definition is valid directly in the correlated electron Hilbert space and retains both the diagonal population contributions and the contributions mediated by inter-eigenstate coherences.

Because the matrix elements $\langle\beta|\hat{n}_i|\alpha\rangle$ are generally nonzero for $\alpha \neq \beta$, $P_i^{orb}(t)$ contains both eigenstate populations and inter-eigenstate coherence contributions. An increase in $P_i^{\mathrm{orb}}$ therefore does not necessarily represent a direct population transfer into localized orbital $i$; it may also arise from coherent redistribution of orbital character within the correlated eigenstates. This distinction is essential for identifying relaxation pathways that are not apparent from populations alone. To resolve directional orbital redistribution, a channel-resolved transfer contribution $I_{i\to j}(t)$ is extracted from the dissipative part of the Redfield evolution. This decomposition assigns the dissipative change of the orbital occupation to directed orbital channels according to the orbital character of the correlated electronic transitions. The resulting $I_{i\to j}(t)$ is a diagnostic constructed from the Redfield dynamics rather than an independently defined microscopic observable. Its directional decomposition is therefore conditional on the adopted orbital-channel partitioning. The cumulative directional transfer over the relaxation interval is then defined as:

$$\Phi_{i\to j} = \int_0^{t_{\mathrm{f}}} P_i^{\mathrm{orb}}(t)\, I_{i\to j}(t)\, dt, \tag{17}$$

where $t_{\mathrm{f}}$ is chosen to encompass the nonequilibrium relaxation interval of interest. The channel decomposition is not unique in the presence of coherent orbital mixing; it is used here as a reproducible diagnostic for comparing directional contributions generated by the same underlying Redfield evolution. The resulting dimensionless quantity is interpreted as an integrated directional transfer weight rather than a directional transfer rate. It quantifies the cumulative contribution of the selected orbital channel over the nonequilibrium relaxation interval. Because $\Phi_{i\to j}$ is obtained from the microscopic Redfield dynamics, it is not an independently fitted parameter. Energy transfer is not inferred from $\Phi_{i\to j}$ alone; the latter characterizes the redistribution of orbital occupation, whereas an energy-resolved transfer measure would additionally require weighting the corresponding microscopic channels by the associated electronic energy changes. The resulting formulation preserves a clear distinction between coherent electronic connectivity and irreversible relaxation. Hybridization, hopping, Coulomb interactions, and spin–orbit coupling determine the correlated spin–orbital manifold and its coherent connectivity, whereas the electron-phonon interaction generates the dissipative kernel responsible for irreversible redistribution. Their combined action determines how a photoexcited carrier redistributes energy and orbital character through intermediate electronic states. Importantly, because the complete non-secular density matrix is retained, pathways with substantial coherence-mediated dynamical contributions can be resolved even when they do not appear as simple direct population transfers. These microscopic fluxes then provide the dynamical quantities against which the spectroscopic visibility of the same pathways is evaluated in the subsequent formulation.

### 2.3 Connection to Multidimensional Spectroscopy and Hidden-Pathway Topology

The microscopic orbital dynamics described by the Redfield quantum kinetics become experimentally accessible through the electronic transition-dipole operator:

$$\hat{\boldsymbol{\mu}} = \sum_{\alpha,\beta} \boldsymbol{\mu}_{\alpha\beta} |\alpha\rangle\langle\beta|, \qquad \boldsymbol{\mu}_{\alpha\beta} = \langle\alpha|\hat{\boldsymbol{\mu}}|\beta\rangle, \tag{18}$$

where $|\alpha\rangle$ and $|\beta\rangle$ are the correlated spin-orbital eigenstates of $H_{\mathrm{el}}$. The magnitude of $\boldsymbol{\mu}_{\alpha\beta}$ determines the optical coupling strength, whereas its vector character specifies the polarization dependence of the transition. Because the correlated eigenstates generally contain contributions from multiple localized orbital and spin configurations, the transition-dipole matrix elements provide the microscopic projection of the correlated electronic dynamics onto the experimentally accessible optical response. The nonlinear optical response is evaluated using the same Liouville-space dynamics that govern the nonequilibrium density matrix. Denoting the Liouvillian associated with the coherent electronic Hamiltonian and the non-secular Redfield dissipator by $\mathcal{L}$, the corresponding propagator is $\mathcal{G}(t) = \exp(\mathcal{L}t)$. The dissipative part of $\mathcal{L}$ is therefore determined by the same electron-phonon correlation functions and Redfield tensor introduced above. The third-order polarization can consequently be expressed in the causal response-function form:

$$P^{(3)}(t) = \int_0^\infty dt_3 \int_0^\infty dt_1\, R^{(3)}(t_3, T_\mathrm{w}, t_1) E(t - t_3) E(t - t_3 - T_\mathrm{w}) E(t - t_3 - T_\mathrm{w} - t_1) \quad (19)$$

where $t_1$ is the initial coherence period, $T_\mathrm{w}$ is the waiting period during which populations and coherences evolve, and $t_3$ is the detection period. The response function $R^{(3)}$ is generated by successive light–matter interactions separated by propagation under $\mathcal{G}(t)$. Thus, the nonlinear optical response and the nonequilibrium quantum kineticss are two manifestations of the same microscopic Hamiltonian and dissipative dynamics rather than independent phenomenological descriptions. For an optical field polarized along the unit vector $\mathbf{e}_\ell$, the relevant dipole operator is $\hat{\mu}_\ell = \mathbf{e}_\ell \cdot \hat{\boldsymbol{\mu}}$. Its action in Liouville space is represented by the dipole superoperator $\mathcal{M}_\ell[\rho] = [\hat{\mu}_\ell, \rho]$, with the overall sign determined by the convention adopted for the light-matter interaction. The third-order response can then be resolved into the conventional Liouville-space contributions $R^{(3)} = R_\mathrm{GSB} + R_\mathrm{SE} + R_\mathrm{ESA}$, corresponding to ground-state bleaching (GSB), stimulated emission (SE), and excited-state absorption (ESA). These contributions represent distinc classes of optical Liouville-space pathways and should not be identified one-to-one with individual microscopic relaxation channels. Their relative amplitudes and waiting-time dependence reflect the evolution of the ground- and excited-state populations and coherences under the same Hamiltonian and Redfield dynamics. In particular, ESA can provide sensitivity to higher-lying excited configurations and intermediate electronic states, thereby supplying complementary information on multistep relaxation pathways. The model spectroscopic signal is obtained by Fourier transformation with respect to the coherence and detection periods,

$$S(\omega_3, T_\mathrm{w}, \omega_1) = \mathrm{Re}\left[\int_0^\infty dt_1 \int_0^\infty dt_3\, R^{(3)}(t_3, T_\mathrm{w}, t_1) e^{-i\omega_1 t_1} e^{+i\omega_3 t_3}\right], \quad (20)$$

where $\omega_1$ and $\omega_3$ denote the excitation and detection frequencies, respectively. Diagonal features predominantly reflect optically allowed transition energies, whereas off-diagonal cross peaks indicate spectroscopic connectivity or dynamical communication between distinct transitions. Their waiting-time dependence can contain signatures of population transfer, coherence transfer, and involvement of intermediate electronic configurations. A cross peak is therefore interpreted here as evidence of spectroscopic connectivity rather than as unique proof of a particular microscopic orbital mechanism. The spectroscopic quantity used in the present analysis is a model spectroscopic proxy constructed from the calculated nonequilibrium populations, inter-state coherences, and projected transition dipoles. It is not intended to represent a full third-order two-dimensional nonlinear spectrum and therefore does not involve an explicit decomposition into rephasing and non-rephasing contributions or into ground-state bleaching, stimulated emission, and excited-state absorption pathways. Instead, the calculated signal provides a consistent projection of the evolving quantum-kinetic dynamics onto transition-energy space and is used to compare the relative spectral visibility of orbital-transfer channels. Assignment of a hidden pathway requires consistency between its spectral evolution, the transition-dipole matrix elements, and the orbital-resolved quantum kineticss generated by the Redfield dynamics. The orbital-resolved quantities used below are evaluated from the occupation-operator definition given in Eq. (16). This quantity contains contributions from both eigenstate populations and inter-eigenstate coherences and thus represents the instantaneous orbital weight generated by the full non-secular Redfield dynamics. The corresponding orbital-state index is defined as:

$$\mathrm{OSI}_i(t) = \frac{P_i^{\mathrm{orb}}(t)}{\sum_{j\in\mathcal{S}} P_j^{\mathrm{orb}}(t)}, \tag{21}$$

When the tracked orbital manifold contains the complete electronic occupation represented in the analysis, $\sum_i P_i^{orb}$ equals the corresponding conserved electron number. The normalized orbital-state index therefore represents the fractional occupation of orbital sector i within the tracked manifold. Here $\mathcal{S}$ denotes the localized-orbital manifold retained in the analysis. When this manifold contains the complete tracked population, $0 \leq \mathrm{OSI}_i(t) \leq 1$ and $\sum_{i\in\mathcal{S}} \mathrm{OSI}_i(t) = 1$. OSI therefore quantifies the instantaneous fractional orbital weight within the selected manifold, but does not by itself distinguish irreversible population transfer from coherence-induced orbital mixing. The persistence of a selected electronic coherence is characterized by the dimensionless coherence-survival factor:

$$\mathrm{CSF}_{\alpha\beta}(t) = \frac{|\rho_{\alpha\beta}(t)|}{|\rho_{\alpha\beta}(t_0)| + \varepsilon_\rho}, \tag{22}$$

where $t_0$ specifies the reference time at which the coherence is established and $\varepsilon_\rho > 0$ is a numerical regularization parameter used only when the reference coherence is vanishingly small. CSF quantifies the persistence of the selected off-diagonal density-matrix element relative to its reference magnitude. It is therefore a derived diagnostic of the Redfield dynamics rather than an independent dynamical variable. Its temporal evolution reflects the combined effects of dissipative decoherence, population relaxation, and coherence transfer contained in the non-secular quantum-kinetic evolution.

The present pathway-resolved quantum-kinetic framework extends conventional descriptions of non-thermal carrier relaxation by resolving orbital populations, inter-orbital energy transfer, quantum coherence, and spectroscopic visibility within a single dynamical representation. Unlike Boltzmann and multi-temperature models, which primarily track carrier distributions [31, 32], the density-matrix formalism retains inter-orbital coherences and permits the separation of population driven and coherence-mediated transfer channels [33–35]. Semiconductor Bloch approaches furnish a microscopic account of light–matter coupling, yet typically introduce relaxation through phenomenological dephasing times [36–38]; the present treatment instead incorporates dissipation via a non-secular Redfield tensor that evolves populations and coherences under the same system bath interaction [39–41]. By applying an identical computational framework to a family of MXenes, the approach enables direct comparison of pathway hierarchies and distinguishes dynamically significant inter-orbital channels from those with strong spectroscopic visibility, thereby resolving the non-equivalence of directional transfer, coherence, and optical response [42–44].

## III. Results & Discussion

To determine how material-specific electronic structure governs nonequilibrium quantum dynamics, four representative MXenes are considered: $Ti_3C_2O_2$, $Ti_3C_2F_2$, $Ti_3C_2(OH)_2$, and $Mo_2TiC_2O_2$. The first three isolate the effect of surface termination within the Ti-based family, whereas $Mo_2TiC_2O_2$ introduces transition-metal substitution. The corresponding structural, electronic, hopping, Coulomb and spin-orbit parameters are given in the Supplementary Information (Tables S1-S4) and are included in the same correlated spin-orbital Hamiltonian without material-specific phenomenological relaxation parameters. Surface termination influences the local electrostatic potential, crystal field, orbital energies and hybridisation, while the transition metal substitution additionally impacts the orbital character, the scales of the interactions, the electronic dispersion and the spin-orbit mixing [45–52]. These material-dependent modifications enter the correlated eigenstates and electron–phonon couplings of the present framework, thereby determining the material-specific non-secular Redfield dynamics and the resulting orbital-resolved response. Experimentally, MXenes exhibit distinct non-equilibrium electronic and phononic energy-transfer channels whose relaxation pathways depend sensitively on electronic structure and electron–phonon coupling [32, 42, 43, 53]. Thus, the material dependence is propagated consistently from the microscopic electronic structure through quantum kinetics to experimentally accessible spectroscopic observables.. Because the same microscopic framework is applied to all four MXenes, differences in orbital-selective relaxation, coherence evolution, transfer flux, and hidden-pathway visibility can be directly traced to material-specific modifications of the electronic structure and its coupling to the phonon environment.

The numerical workflow begins from a material-specific five-orbital spin-resolved Hamiltonian that incorporates crystal-field energies, orbital-resolved hopping and hybridisation amplitudes, atomic spin–orbit coupling, and projected dipole matrix elements. After a single diagonalisation, the eigenbasis is retained for all subsequent operations. Coherent optical excitation is introduced through the electric-dipole interaction with an 800 nm transform-limited Gaussian pulse of 30 fs duration and peak intensity $3.3 \times 10^{12}$ W cm$^{-2}$. The reduced density matrix is initialized as the projector onto the lowest-energy correlated electronic eigenstate and propagated for 1.5 ps on a uniform 2 fs grid by piecewise exact exponentiation of the time-dependent Liouvillian. This pure-state initialization corresponds to a fixed-electron-number correlated ground state rather than a grand-canonical Fermi–Dirac distribution. The Liouvillian combines unitary Hamiltonian evolution with a Born–Markov non-secular Redfield superoperator generated from an Ohmic phonon bath with coupling strength 0.035 and cut-off frequency 1.0 eV at 300 K. Hermiticity and unit trace are monitored at every time step, and no positivity correction is applied. Projection of the instantaneous density matrix onto the original spin-resolved orbital subspaces yields the time-dependent orbital populations. The Redfield-induced orbital population changes are partitioned into population-only and coherence-mediated components, furnishing orbital-resolved coherence weights. These populations and coherence weights, together with the projected inter-orbital dipole couplings, define approximate diagonal-peak and cross-peak intensities. Hidden-pathway visibility is obtained as the ratio of each cross-peak intensity to the sum of the corresponding diagonal intensities. A dynamical-weight matrix constructed from orbital population differences and coherence fractions is combined with the normalised visibility matrix to produce the directed transfer topology that ranks the microscopic inter-orbital pathways. The identical numerical protocol, including grid densities, bath parameters, pulse characteristics, and pathway definitions, is applied to all four MXenes; only the material-specific microscopic parameter sets listed in Tables S1–S4 are varied. Systematic sweeps of numerical and physical parameters assess the robustness of the resulting spectral features, pathway hierarchies, and derived observables across the explored parameter ranges.

The bath spectral environment together with the resulting orbital-resolved quantum kinetics is depicted in Figure 2. Panel (a) shows a bath spectral density that is peaked at low excitation energies and decays toward higher energies, thereby defining the frequency-dependent weighting with which the environment enters the dissipative kernel. This spectral profile sets the energy window over which bath-induced transitions and pure-dephasing processes contribute to the reduced system dynamics. Panels (b–e) resolve the population dynamics of two distinct $d$-orbital channels, $d_{xy}$ and $d_{x^2-y^2}$, while $C_{\mathrm{tot}}$ (secondary axis) quantifies the total coherence. The two orbital populations evolve along non-identical trajectories, reflecting their unequal coupling strengths to the quantum-kinetic relaxation channels. The contrast between solid and dashed curves isolates the contribution of population–coherence coupling that is retained in the non-secular treatment but eliminated by secularization. In $Ti_3C_2O_2$ the $d_{xy}$ population exhibits a transient rise followed by depletion, whereas the $d_{x^2-y^2}$ population remains substantially occupied throughout. $Ti_3C_2F_2$ displays a more pronounced late-time suppression of the $d_{xy}$ population together with a persistent $d_{x^2-y^2}$ component. $Ti_3C_2(OH)_2$ retains a sizable $d_{x^2-y^2}$ population while the $d_{xy}$ channel decays at longer times. $Mo_2TiC_2O_2$ shows both the largest coherence amplitude and the strongest late-time variation in $C_{\mathrm{tot}}$, accompanied by substantial redistribution between the two orbital populations. The material-dependent separation between non-secular and secular trajectories originates from the structure of the Redfield kernel rather than from the population equations alone. Panel (f) reveals finite off-diagonal population–coherence matrix elements within the retained quantum manifold, demonstrating that the population and coherence sectors remain kinetically coupled. These terms allow coherence-dependent components of the density matrix to influence the population evolution. Their removal under the secular approximation alters the accessible relaxation pathways and accounts for the differences between the solid and dashed curves in (b–e).

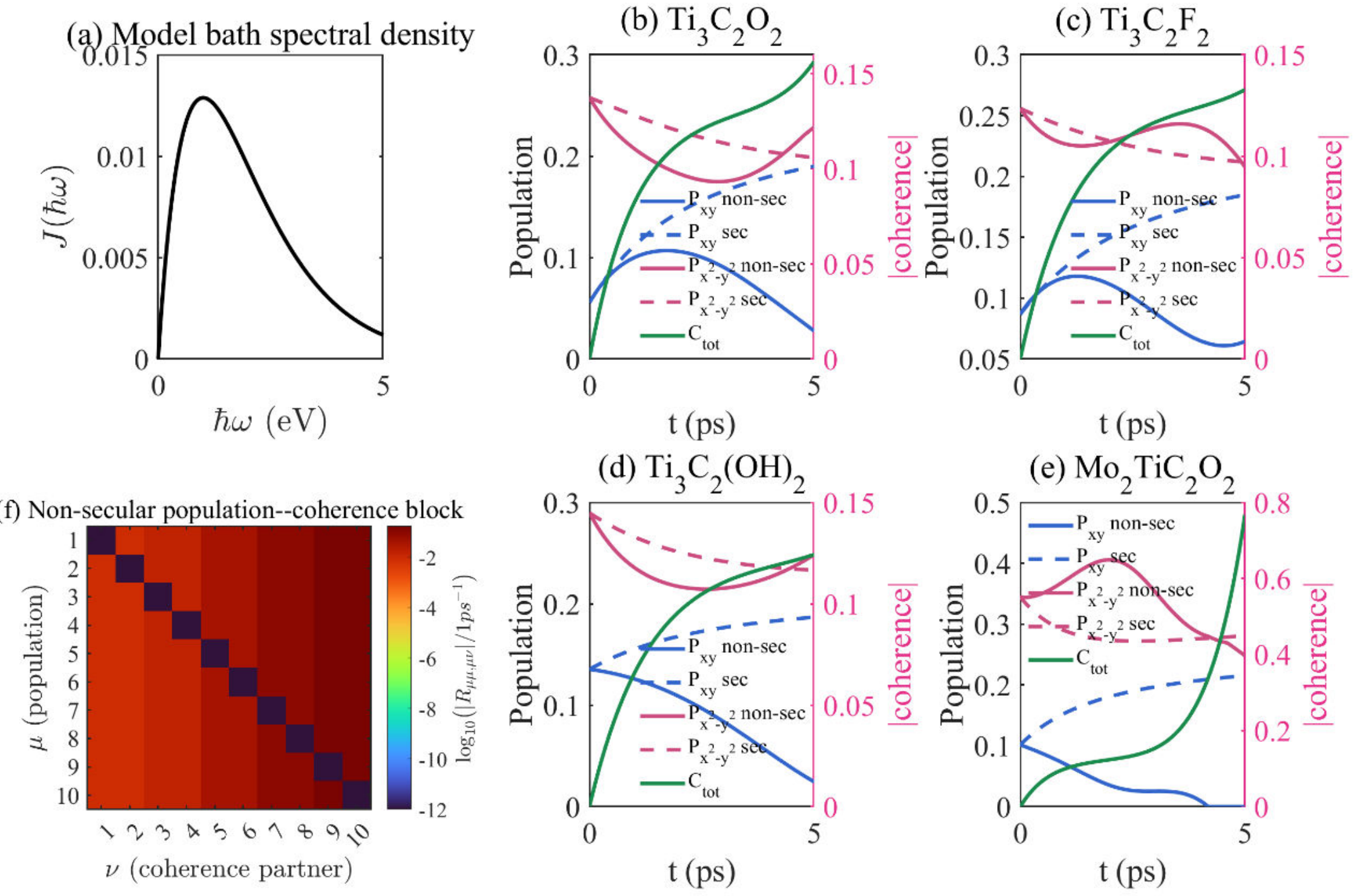

**Figure 2:** Bath spectrum and material-dependent orbital dynamics. (a) Model bath spectral density . (b–e) Time-dependent populations of the and orbitals and total coherence for the four MXenes, comparing non-secular and secular quantum-kinetic dynamics. (f) Logarithmic magnitude of the non-secular population–coherence Redfield block.

The six panels in Figure 3 resolve how non-thermal carrier relaxation is distributed across orbital populations, energy, orbital selectivity, and coherence. Panels (a,b) show strongly non-uniform relaxation within the d manifold. In $Ti_3C_2O_2$, population changes extend across several orbitals, with $d_{z^2}$ remaining dominant while $d_{xz}$ increases and $d_{yz}$ decreases. $Mo_2TiC_2O_2$ exhibits a stronger redistribution, where late-time growth of $d_{yz}$ accompanies depletion of $d_{xy}$, $d_{xz}$, and $d_{z^2}$. These trajectories reveal an orbital-dependent relaxation pathway that cannot be represented by a single population coordinate. Panels (c,d) provide complementary measures of the evolving carrier distribution. The three Ti-based systems retain similar DNT values and a narrow range of excess energies. $Mo_2TiC_2O_2$ starts from the largest excess energy and develops pronounced late-time reductions in both Eexcess and DNT, consistent with a substantial modification of its non-thermal carrier distribution. The orbital-selectivity measure in (e) increases at late times, most strongly for $Mo_2TiC_2O_2$, while the excess energy continues to decrease. The relaxation of carrier energy is thus accompanied by an increasingly uneven distribution among orbital channels, showing that orbital differentiation persists during energy relaxation. Coherence dynamics in (f) remain comparatively weak in the Ti-based systems but increase strongly at late times in $Mo_2TiC_2O_2$. The simultaneous enhancement of orbital selectivity and coherence, together with the pronounced redistribution within the d manifold, distinguishes its relaxation dynamics from those of the Ti-based MXenes. Population redistribution, orbital differentiation, and coherence evolve on different temporal trajectories, indicating that the non-thermal relaxation pathway retains distinct orbital and coherence-dependent components.

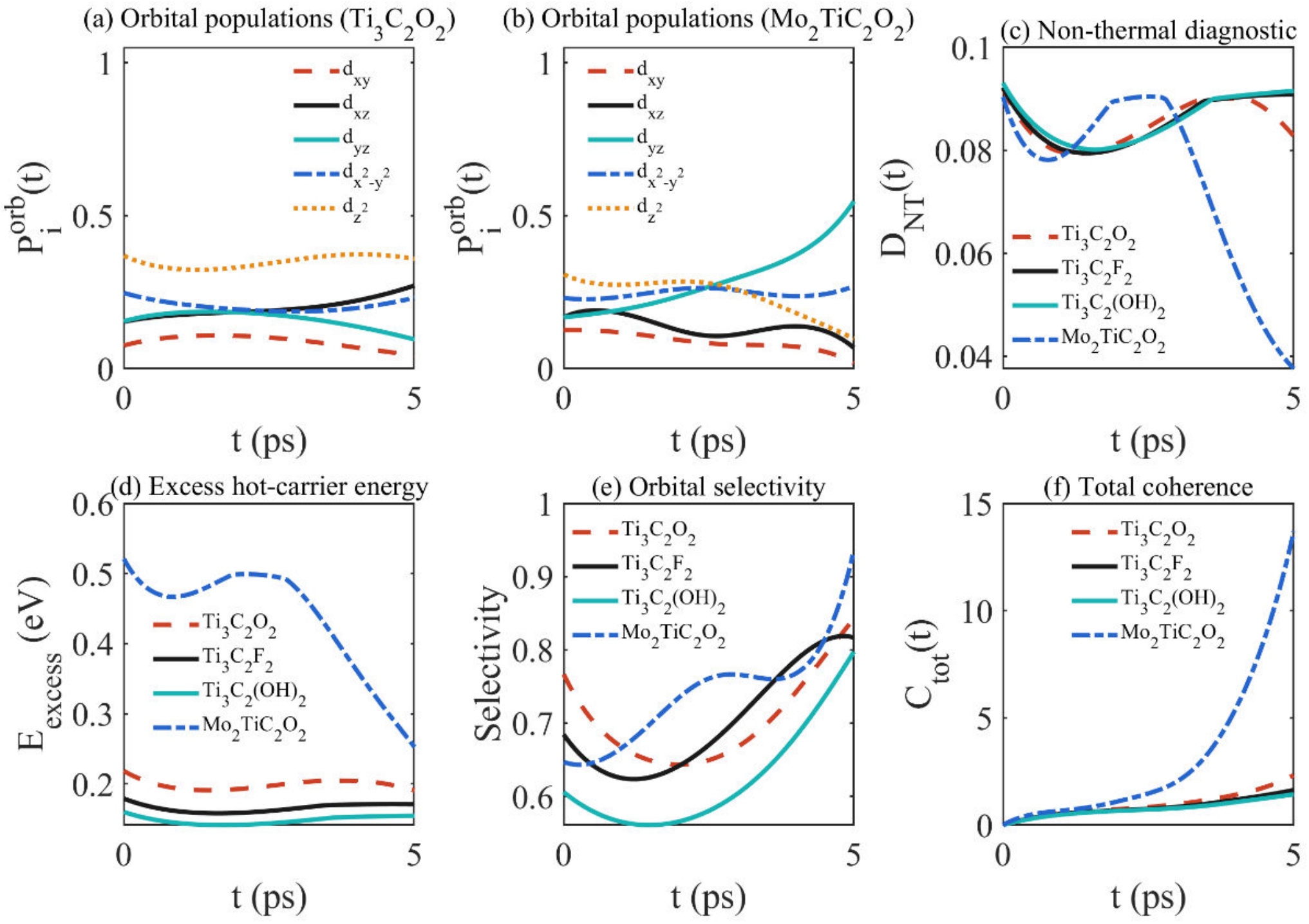

**Figure 3:** Orbital-resolved signatures of non-thermal carrier relaxation. (a,b) Time-dependent populations of the five -orbitals in TiCO and MoTiCO. (c–f) Evolution of the non-thermal diagnostic , excess hot-carrier energy , orbital selectivity, and total coherence for the four MXenes.

The population dynamics in panels (a,b) are directly related to the time-integrated inter-orbital fluxes shown in panels (c,d) in Figure 4. In $Ti_3C_2O_2$, the increase of the $d_{xz}$ population matches the largest cumulative contribution from the $d_{yz} \rightarrow d_{xz}$ channel. The simultaneous decrease of $d_{yz}$ is distributed across several outgoing pathways, principally $d_{yz} \rightarrow d_{x^2-y^2}$ and $d_{yz} \rightarrow d_{z^2}$. Because the integrated flux is shared among multiple channels, the net redistribution remains comparatively modest, consistent with the similar pathway weights displayed in panel (e). The early rise and subsequent flattening of several cumulative-flux curves further indicate that the underlying inter-orbital transfer rates decrease as the populations evolve toward a more slowly varying distribution. The $Mo_2TiC_2O_2$ follows a distinctly different redistribution pattern. The pronounced late-time growth of $d_{yz}$ is accompanied by a substantially larger cumulative influx from $d_{z^2} \rightarrow d_{yz}$, together with secondary contributions from $d_{xz} \rightarrow d_{yz}$ and $d_{xy} \rightarrow d_{yz}$. The depletion of $d_{z^2}$, $d_{xz}$, and $d_{xy}$ therefore tracks the rise of the $d_{yz}$ population, while the observed population trajectories represent the net outcome of all incoming and outgoing channels. The dominant $d_{z^2} \rightarrow d_{yz}$ pathway continues to accumulate strongly at later times, producing the clear hierarchy of pathway strengths seen in panels (d,f). The larger magnitude of the integrated fluxes and the greater separation between the leading and secondary channels indicate that orbital redistribution in $Mo_2TiC_2O_2$ is concentrated more strongly into the $d_{yz}$ orbital than in $Ti_3C_2O_2$. Comparison of the two materials thus links the observed population changes to their underlying inter-orbital transfer channels and shows that similar d-orbital manifolds can redistribute population through markedly different flux hierarchies.

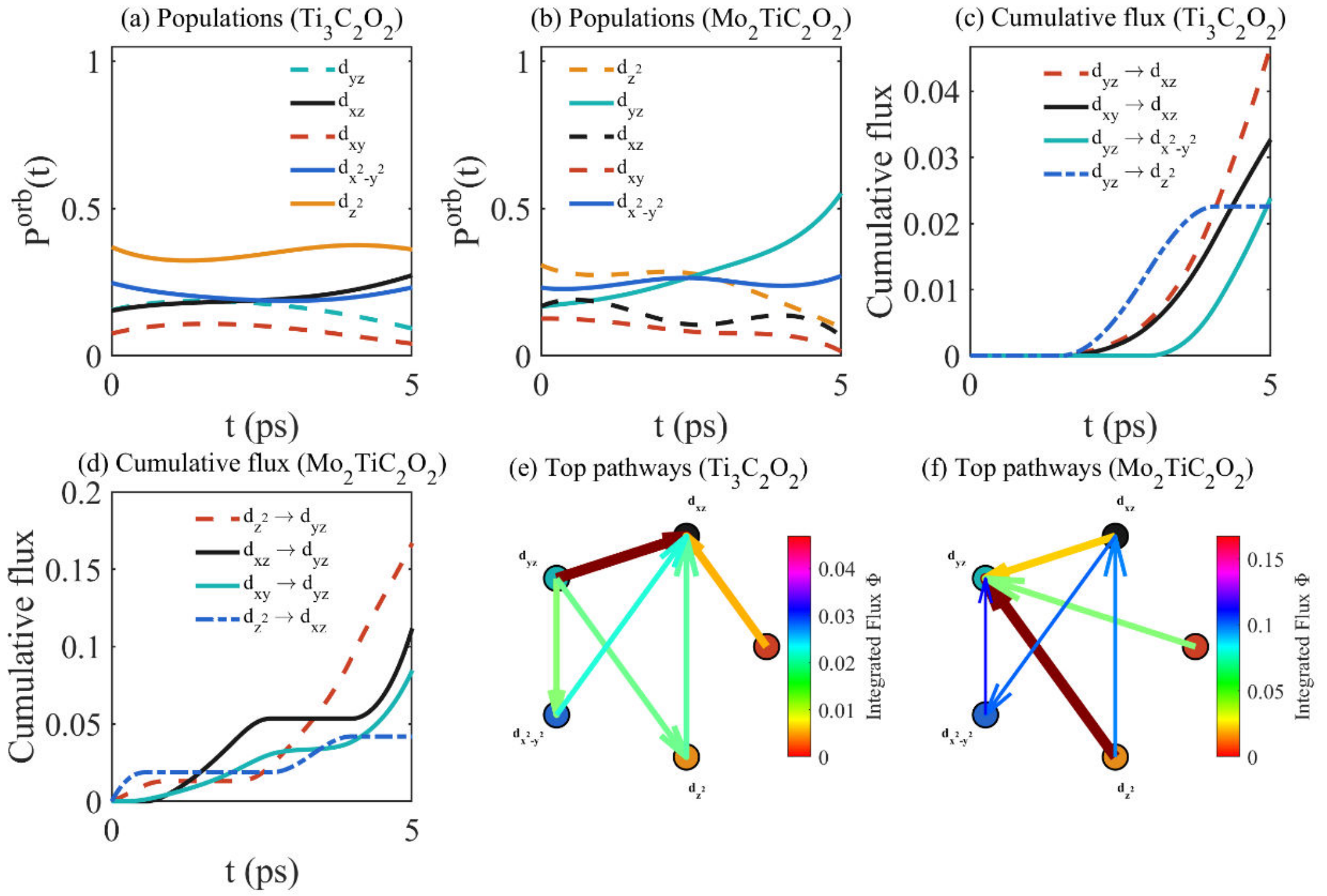

**Figure 4:** Orbital population dynamics and dominant inter-orbital relaxation pathways in $Ti_3C_2O_2$ and $Mo_2TiC_2O_2$. (a,b) Time-dependent populations of the d orbitals. (c,d) Cumulative inter-orbital fluxes along the dominant relaxation channels. (e,f) Network representations of the strongest pathways, with arrow direction indicating the direction of population transfer and color encoding the integrated flux.

The panels (a–c) in Figure 5 show that the hierarchy of the dominant inter-orbital transfer pathways is only weakly modified by temperature. The pathway strengths in panel (a) remain nearly constant across the investigated range and preserve the ordering of the leading channels. The branching ratios in panel (b) exhibit similarly small variations, indicating that thermal changes mainly adjust the relative weights of existing channels rather than reorder the dominant pathways. The hierarchy ratios in panel (c) remain above unity for all materials and display only moderate temperature dependence. $Ti_3C_2F_2$ retains the largest ratio and shows a slight increase at higher temperature, while $Ti_3C_2(OH)_2$ and $Mo_2TiC_2O_2$ undergo gradual reductions. These changes modify the separation between competing channels without substantially altering the underlying pathway hierarchy. Panels (d–f) show the response to the normalized pump amplitude. The leading pathway strengths in panel (d) remain almost invariant over the explored field range, indicating that the same set of inter-orbital channels remains dominant as the excitation strength increases. The branching ratios in panel (e) vary only moderately, with $Ti_3C_2F_2$ showing a weak decrease at larger pump amplitudes. The hierarchy ratios exhibit the clearest pump dependence. $Ti_3C_2F_2$ displays a non-monotonic response, with an enhanced separation between the dominant and competing pathways at intermediate excitation followed by a reduction at larger fields. $Mo_2TiC_2O_2$ shows the opposite late-field trend, with its hierarchy ratio increasing under stronger pumping. Temperature and pump amplitude therefore preserve the identity and ordering of the leading pathways across the investigated ranges while modifying the separation between dominant and secondary transfer channels.

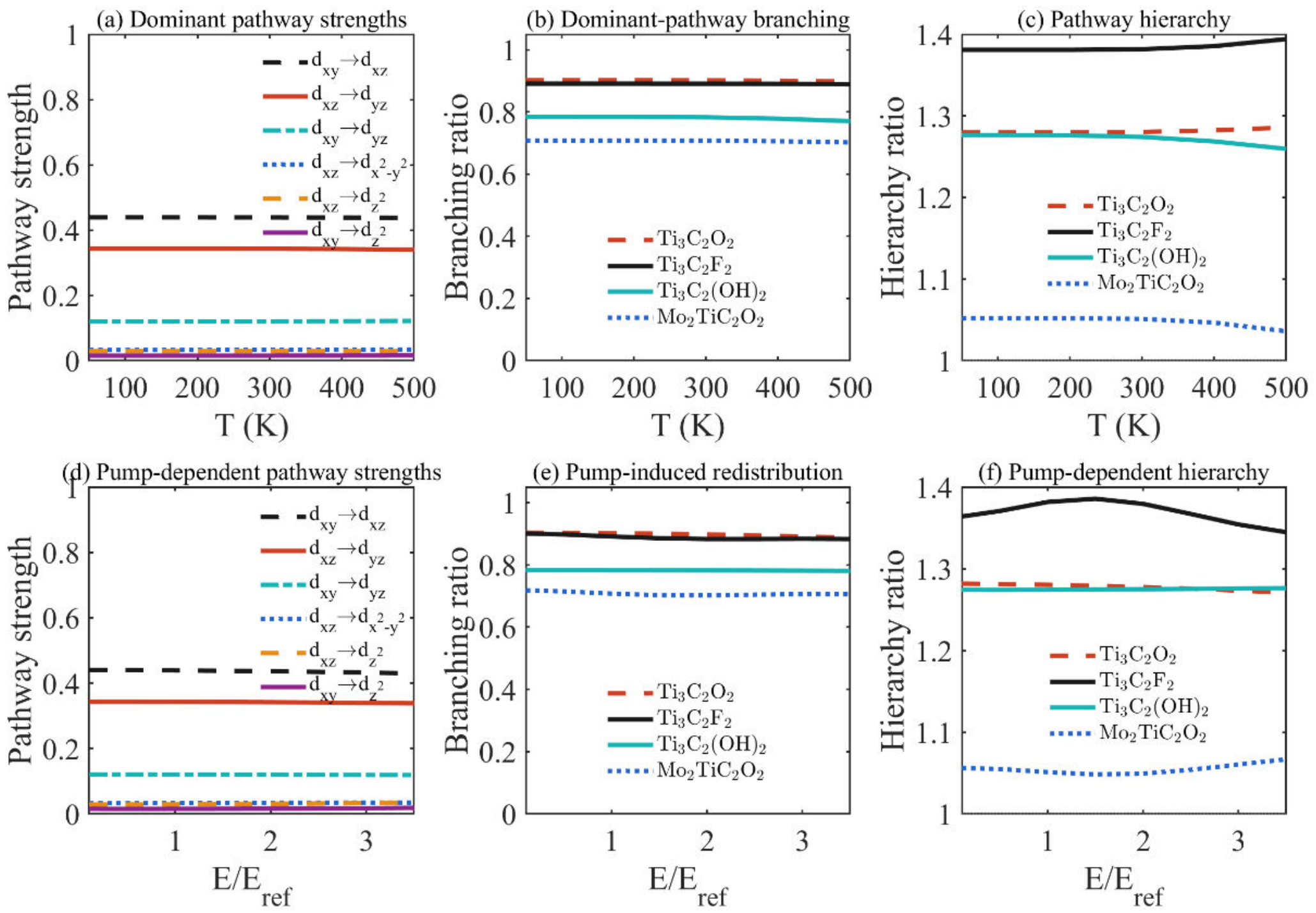

**Figure 5:** Thermal and pump-field dependence of the dominant inter-orbital pathway hierarchy. (a,d) Relative strengths of the leading inter-orbital pathways as functions of temperature and normalized pump amplitude, respectively. (b,e) Branching ratios of the dominant pathway across the investigated MXenes. (c,f) Corresponding pathway hierarchy ratios, comparing the separation between the leading and competing transfer channels.

The differential absorption spectrum responds to variations in the parameters governing thermal occupation, relaxation, secular mixing, non-secular coupling, pure dephasing, and bath-memory regularization in Figure 6. Across all six parameter sweeps, the principal spectral features remain localized within similar phonon-energy intervals, while their amplitudes and relative weights change substantially. The parameter dependence therefore acts primarily on the magnitude and relative contribution of the underlying dissipative channels rather than on large displacements of the main spectral features. Phonon temperature progressively amplifies both positive and negative spectral excursions, consistent with an increased contribution from thermally populated phonon channels. The relaxation-rate sweep produces the largest variation in $\Delta_{Abs}$: the weakly damped case $\gamma_0/\gamma_{ref} = 0.1$ develops substantially larger oscillatory amplitudes than the more strongly damped regimes. The secular-mixing parameter $\lambda$ or $\xi_{sec}$ continuously interpolates between the non-secular and secular Redfield generators. This secular mixing also strongly affects the spectral amplitude; the response is largest near $\lambda = 0$ and becomes progressively attenuated as $\lambda$ approaches the fully secular limit. The subsequent sweeps isolate the sensitivity to non-secular coupling and pure dephasing. Increasing the non-secular coefficient enhances the amplitude of the dominant positive and negative features, whereas changes in the pure-dephasing coefficient reshape their local magnitudes and deepen the negative excursion associated with the higher-energy feature. The soft-cutoff parameter affects the spectral amplitude through the numerical regularization of the bath-correlation transform: shorter cutoff times produce stronger excursions, whereas longer cutoff times yield a more attenuated response. The positions of the spectral features remain comparatively robust across the sweep, while their amplitudes vary with the relative contributions of thermal occupation, relaxation, coherence damping, non-secular mixing, and the chosen numerical regularization. The weak dependence of the calculated dynamics on the soft-cutoff parameter indicates stability against the selected regularization of the bath-correlation transform rather than a sensitivity to a physical bath-memory timescale.

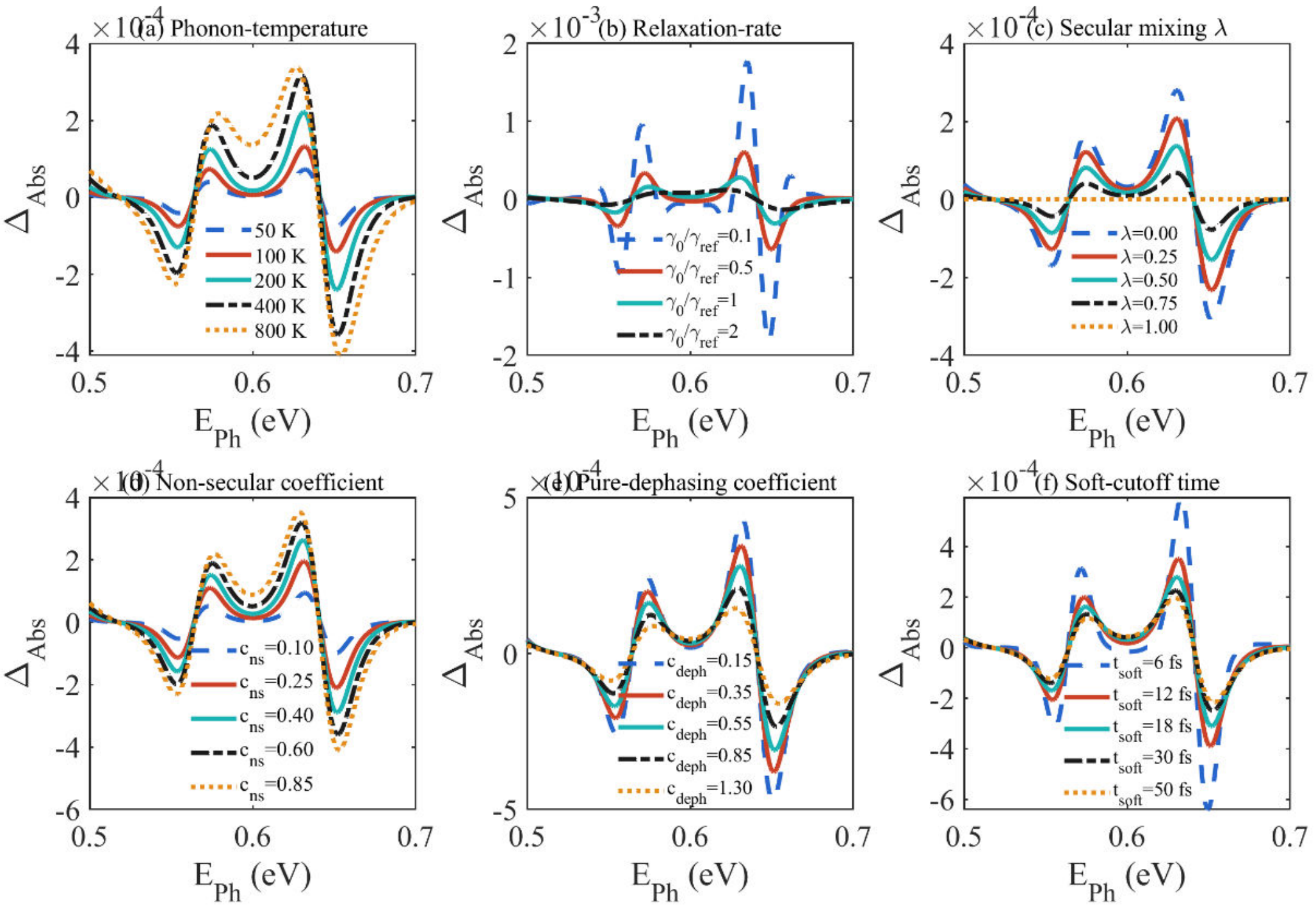

**Figure 6:** Sensitivity of the phonon-energy-dependent differential absorption to open-system parameters. $\Delta_{Abs}$ is shown as a function of phonon energy $E_{ph}$ for variations in (a) phonon temperature, (b) relaxation rate, (c) secular mixing parameter $\lambda$ or $\xi_{sec}$, (d) non-secular coefficient, (e) pure-dephasing coefficient, and (f) Sensitivity to the numerical soft-cutoff time used in the bath-correlation transform.

The time-dependent redistribution of orbital populations is linked to the accumulation of inter-orbital flux and the resulting spectroscopic pathway visibility in Figure 7. The initially dominant dxy population decreases rapidly, while the $d_{x^2-y^2}$ and $d_{z^2}$ populations increase and the $d_{yz}$ and $d_{xz}$ populations undergo more moderate redistribution. To compare dynamical importance with spectroscopic accessibility, the calculated pathway quantities are further analyzed using derived post-processing metrics. The hidden-pathway analysis does not introduce an additional dynamical equation or modify the Redfield evolution. Instead, it combines the directional transfer contribution of a given orbital channel with its corresponding spectroscopic visibility to identify channels whose dynamical contribution is appreciable while their representation in the model spectroscopic proxy is comparatively weak. The resulting hidden-pathway quantities are therefore diagnostic measures of the calculated dynamics rather than independent microscopic observables or transition probabilities. The hidden-pathway visibility (HPV) is used here as a relative visibility metric obtained from the same spectral representation used for the corresponding orbital channel. It quantifies how strongly a dynamically identified pathway is represented in the calculated spectroscopic proxy relative to the selected reference contribution. HPV is therefore a comparative spectroscopic measure and should not be interpreted as a transition probability, population fraction, or independent microscopic rate.

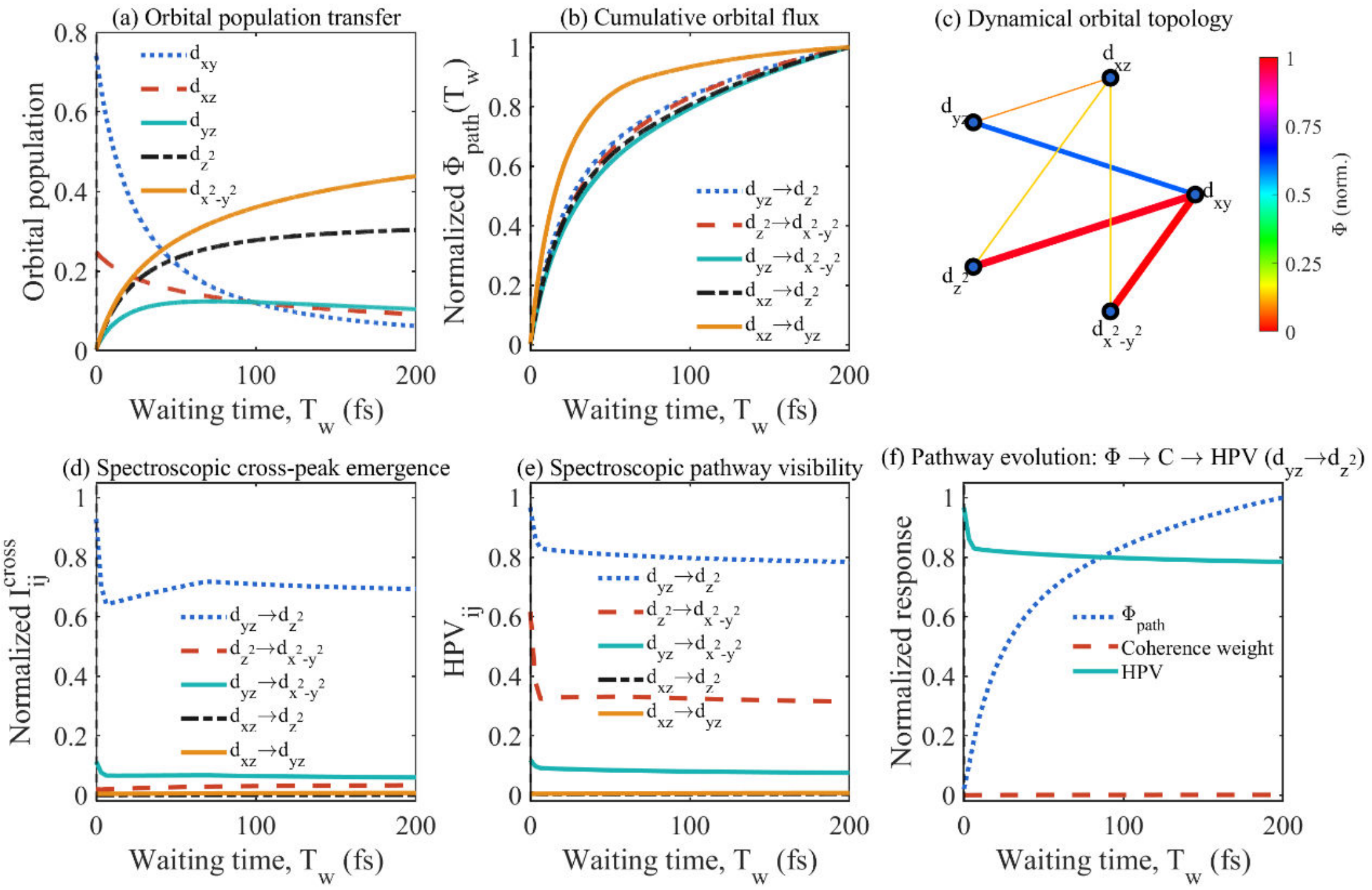

**Figure 7:** Dynamical evolution of orbital transfer pathways and their spectroscopic visibility. (a) Time-dependent populations of the d orbitals. (b) Normalized cumulative fluxes along the leading inter-orbital pathways. (c) Dynamical orbital topology, with edge width and color representing pathway strength and normalized integrated flux, respectively. (d) Emergence of normalized cross-peak features in the model spectroscopic proxy, (e) Time evolution of the hidden-pathway visibility (HPV) for the corresponding pathways. (f) Evolution of the normalized pathway flux, coherence weight, and HPV for the $d_{yz} \rightarrow d_{z^2}$ channel.

Furthermore, the corresponding cumulative fluxes reveal that the leading pathways do not accumulate at the same rate. The $d_{xz} \rightarrow d_{yz}$ channel approaches its final integrated value most rapidly, whereas the $d_{xz} \rightarrow d_{z^2}$ pathway exhibits a slower buildup. The normalized cumulative fluxes are referenced to the largest integrated channel over the selected pathway set, so that their long-time values retain the relative hierarchy of the accumulated transfer strengths. The normalization therefore preserves comparisons between pathways while placing the dominant channel on a dimensionless scale. The pathway hierarchy in orbital space is resolved by the network representation. The strongest connections link $d_{z^2}$ with dxy and $d_{x^2-y^2}$, while the remaining pathways provide weaker links between the orbital nodes. The network therefore describes a structured distribution of integrated transfer strengths rather than a uniform redistribution across the d manifold. The coexistence of several pathways is consistent with the population trajectories, each of which reflects the net balance of multiple incoming and outgoing transfer channels. The spectroscopic observables display a hierarchy distinct from that of the normalized cumulative fluxes. The $d_{yz} \rightarrow d_{z^2}$ cross-peak remains the strongest feature, whereas several pathways that carry finite cumulative flux generate substantially weaker cross-peak signals. The hierarchical pathway visibility preserves this separation: the $d_{yz} \rightarrow d_{z^2}$ pathway maintains the largest visibility, while the other channels remain considerably smaller. The spectroscopic prominence of a pathway is therefore not determined solely by its accumulated population flux but also depends on the pathway-specific factors entering the cross-peak response. The $d_{yz} \rightarrow d_{z^2}$ channel is followed directly in panel (f). Its normalized cumulative flux increases continuously with waiting time, while the coherence weight remains close to zero on the displayed scale. The hierarchical pathway visibility remains high after a rapid initial adjustment and decreases only gradually as the pathway flux continues to accumulate. The distinct temporal profiles of pathway flux, coherence weight, and hierarchical pathway visibility show that the accumulation of population transfer and the evolution of spectroscopic visibility are not proportional over the displayed waiting-time range.

For the same channels, the hidden-pathway quantities are evaluated as derived combinations of the calculated directional transfer and spectroscopic visibility, allowing dynamically significant but comparatively weakly represented channels to be identified. The selected inter-orbital pathways exhibit strongly time-dependent directional transferes, with distinct amplitudes and oscillation patterns across the $d_{xz} \rightarrow d_{z^2}$, $d_{xy} \rightarrow d_{xz}$, and $d_{xz} \rightarrow d_{x^2-y^2}$ channels in Figure 8. These differences are not reproduced proportionally in the spectroscopic visibilities. Pathways with comparable or overlapping flux amplitudes can display different visibility profiles, while pronounced flux excursions do not always coincide with maxima of the spectroscopic proxy. The time-resolved spectroscopic signal therefore does not furnish a direct one-to-one measure of instantaneous pathway flux. The derived hidden-pathway metrics introduce a further hierarchy that differs from both the directional transfer and spectroscopic-visibility patterns. The three pathways exchange their relative prominence throughout the displayed time window, with repeated changes in their instantaneous strengths. Pathway ranking therefore depends on the metric used to resolve the underlying orbital dynamics. The same inter-orbital transfer can occupy different positions within the flux, spectroscopic, and hidden-pathway hierarchies. The collective contribution of the hidden sector is resolved separately. The normalized hidden-pathway response and ηhidden remain close to unity with relatively weak temporal modulation, whereas the normalized total-flux response stays substantially smaller and oscillates more strongly. The contrast indicates that the hidden-pathway response remains comparatively stable over the same interval in which the directly resolved total flux undergoes pronounced temporal variation. The leading hidden-topology eigenmode displays large-amplitude oscillations and repeated redistribution between high- and low-amplitude intervals, whereas the $d_{z^2}$ reference mode remains nearly time-independent. The hidden-topology response is thus concentrated in a strongly time-dependent leading mode rather than being distributed uniformly across the displayed eigenmode amplitudes. This response remains present across the four MXenes but exhibits material-dependent temporal modulation. $Ti_3C_2O_2$, $Ti_3C_2F_2$, $Ti_3C_2(OH)_2$, and $Mo_2TiC_2O_2$ remain within a narrow high-response range while differing in the timing and amplitude of their excursions, reflecting differences in the temporal evolution of their hidden-pathway responses.

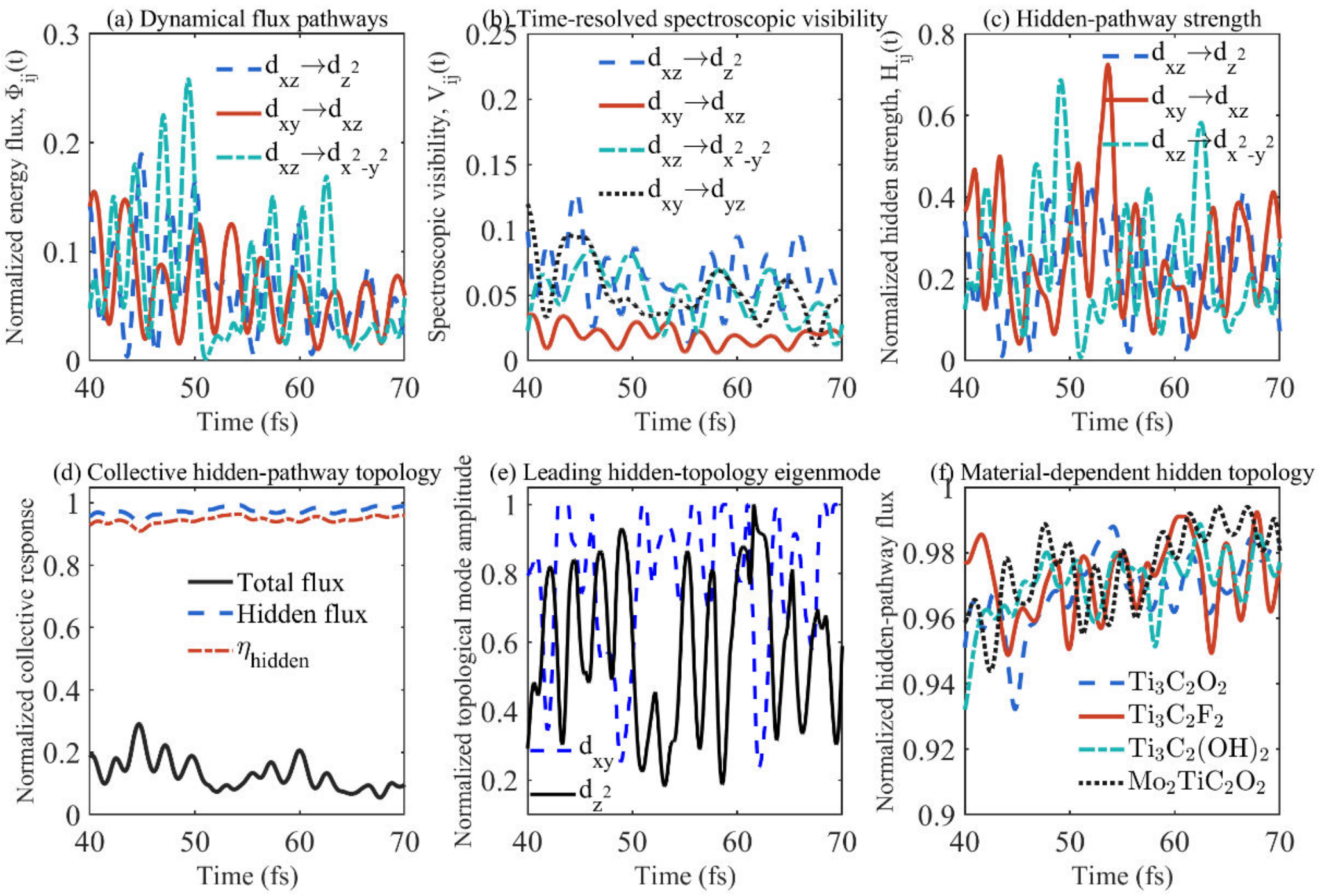

**Figure 8:** Dynamical emergence and material dependence of hidden inter-orbital pathways. (a) Time-dependent normalized directional transferes along selected orbital-transfer channels. (b) Corresponding time-resolved spectroscopic visibilities. (c) Derived hidden-pathway metrics for the same channels. (d) Collective response comparing the total flux, hidden-pathway flux, and hidden-pathway fraction ηhidden. (e) Temporal evolution of the leading hidden-topology eigenmode relative to the $dz^2$ mode. (f) Material dependence of the normalized hidden-pathway flux in $Ti_3C_2O_2$, $Ti_3C_2F_2$, $Ti_3C_2(OH)_2$, and $Mo_2TiC_2O_2$.

The orbital populations undergo pronounced redistribution over the displayed time interval in Figure 9. The $d_{xy}$ population exhibits repeated oscillations with a gradually decreasing amplitude, while the $d_{xz}$ population oscillates out of phase with dxy over much of the time window. The $d_{yz}$ population remains comparatively weakly varying, whereas the $d_{z^2}$ and $d_{x^2-y^2}$ populations evolve more slowly and approach different values near the end of the displayed interval. The redistribution is therefore unevenly distributed across the orbital manifold rather than proceeding through a single common relaxation coordinate. The cumulative orbital energy flow increases monotonically at all temperatures, with the higher-temperature curves remaining above the lower-temperature curves throughout the displayed interval. The temperature dependence is strongest at early and intermediate times and becomes less pronounced toward later times as the normalized cumulative flows approach similar values. The curves indicate temperature-dependent accumulation rates within a common long-time normalized energy-flow range. The hidden-pathway spectral response is concentrated into three narrow features near approximately 0.2, 0.8, and 1.45 eV. Their unequal amplitudes resolve a spectroscopic hierarchy, with the highest-energy feature carrying the largest normalized response and the intermediate-energy feature remaining weaker. The spectral representation therefore isolates discrete transition-energy windows associated with the hidden-pathway response. The hidden-pathway contribution is separated from the total pathway current. Both quantities evolve non-monotonically, while the hidden contribution remains below the total response throughout the displayed time window. Their temporal variations partly overlap, indicating that the hidden component evolves within the broader dynamics of the microscopic current while contributing only a fraction of its normalized response. The hidden-pathway fraction undergoes substantial temporal modulation. Its largest values occur near the central part of the displayed time window, followed by a rapid reduction and a partial recovery at later times. Temperature modifies the magnitude of these excursions without qualitatively changing their temporal structure. The curves remain closely grouped, indicating that the time dependence of ηhidden is more pronounced than its temperature-induced variation. The activation of individual hidden pathways exhibits distinct temperature trends. The $d_{xz} \rightarrow d_{x^2-y^2}$ channel decreases gradually with increasing temperature, whereas the $d_{xz} \rightarrow d_{z^2}$ and $d_{xy} \rightarrow d_{yz}$ channels increase. The $d_{xy} \rightarrow d_{xz}$ pathway remains nearly temperature independent over the explored range. These different trends preserve a pathway-dependent hierarchy while showing that temperature modifies the relative activation of individual hidden channels.

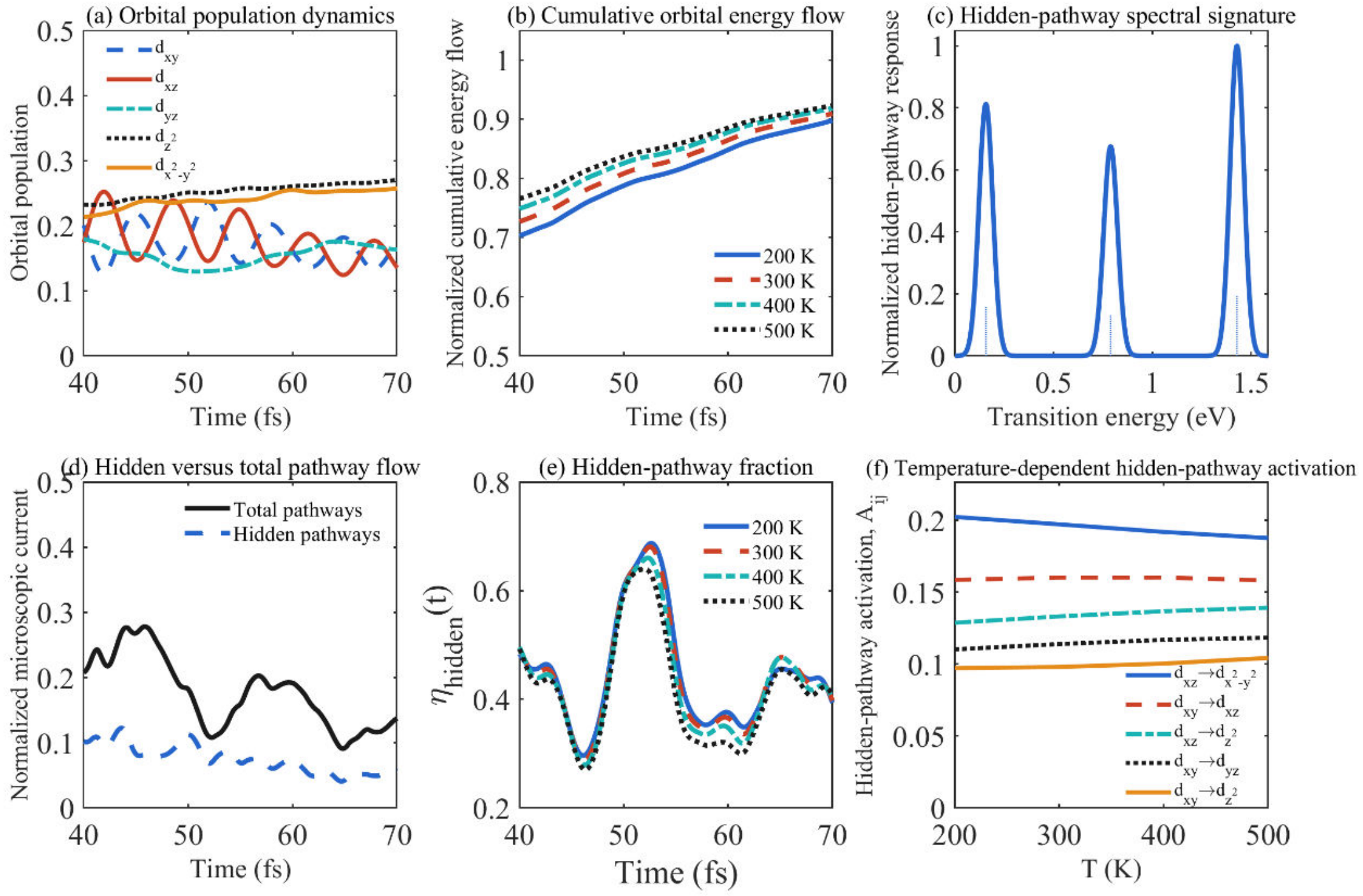

**Figure 9:** Orbital dynamics, spectroscopic signature, and thermal activation of hidden inter-orbital pathways in $Ti_3C_2O_2$. (a) Time-dependent orbital populations. (b) Normalized cumulative orbital energy flow at different temperatures. (c) Normalized hidden-pathway spectral response as a function of transition energy. (d) Comparison between total and hidden pathway contributions to the normalized microscopic current. (e) Time evolution of the hidden-pathway fraction at different temperatures. (f) Temperature dependence of hidden-pathway activation for the leading inter-orbital channels.

The hidden-pathway spectral flow is concentrated within several discrete transition-energy windows whose relative intensities change with time in Figure 10. The low-energy feature near 0.15–0.3 eV dominates at selected times, whereas the spectral weight around approximately 0.6–1.0 eV undergoes substantial temporal redistribution. The higher-energy feature near 1.4–1.5 eV also varies strongly between the displayed time slices. The hidden spectral response therefore evolves mainly through changes in the relative spectral weight of distinct transition-energy channels rather than through continuous displacement of the spectral features. Topological participation is distributed non-uniformly across the d manifold. The $d_{xz}$ orbital retains a comparatively large participation over much of the time window, while the remaining orbitals exhibit oscillatory exchanges in their relative contributions. Energy-weighted participation follows a different ordering and temporal structure. Orbitals with comparable topological participation can carry different energy-weighted contributions, showing that participation in the orbital network and energetic participation constitute distinct measures of the evolving dynamics. The dominant microscopic pathways display strongly pathway-dependent temporal behavior. The $d_{xz} \rightarrow d_{x^2-y^2}$ and $d_{xy} \rightarrow d_{xz}$ channels develop pronounced intermittent maxima, while the other pathways contribute over different temporal intervals and with smaller or more distributed amplitudes. Energy redistribution is therefore not associated with a fixed pathway hierarchy throughout the entire time window. Relaxation-pathway entropy exhibits pronounced temporal modulation, including repeated minima and recoveries, whereas the temperature-dependent curves remain closely grouped. The temporal variation of pathway entropy is larger than the variation induced by temperature over the explored range. Temperature modifies the local magnitude of the entropy without substantially changing the sequence of its principal temporal features. A material-dependent hidden-pathway spectrum is resolved across the four MXenes. All compounds exhibit a strong low-energy response near approximately 0.15–0.2 eV, while the higher-energy structures differ in both position and relative amplitude. $Ti_3C_2F_2$ and $Mo_2TiC_2O_2$ develop comparatively pronounced secondary features, whereas $Ti_3C_2O_2$ and $Ti_3C_2(OH)_2$ display different distributions of spectral weight across the intermediate-energy region. The hidden-pathway response retains a common low-energy spectral feature across the material family, while the higher-energy contributions exhibit material-dependent differences in spectral weight and transition energy. These features should be interpreted as spectral signatures of the derived pathway metrics within the present model, rather than as direct measurements of microscopic hidden-transition probabilities.

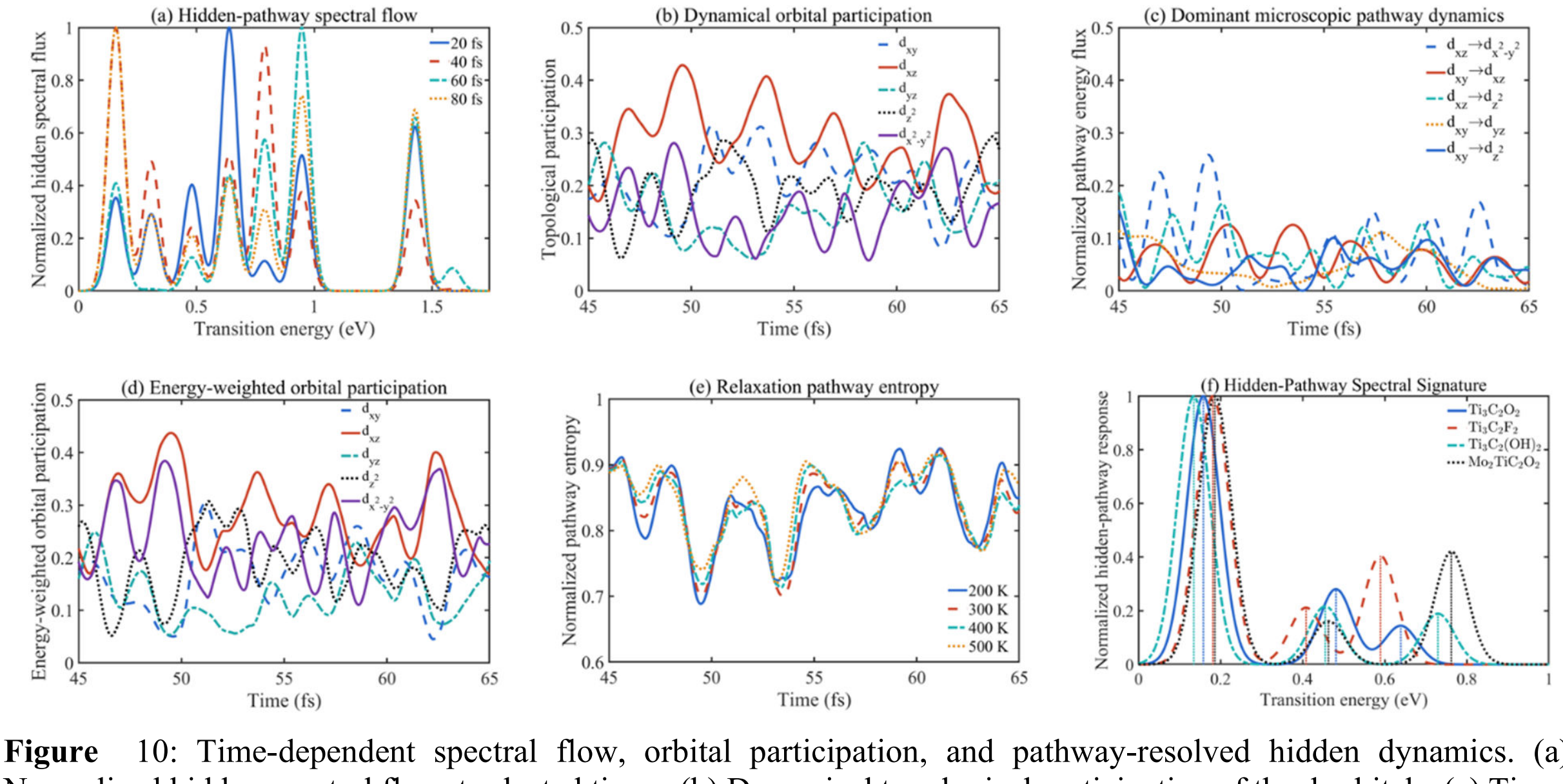

**Figure** 10: Time-dependent spectral flow, orbital participation, and pathway-resolved hidden dynamics. (a) Normalized hidden spectral flux at selected times. (b) Dynamical topological participation of the d orbitals. (c) Time-dependent directional transfer along the dominant microscopic inter-orbital pathways. (d) Corresponding energy-weighted orbital participation. (e) Normalized relaxation-pathway entropy at different temperatures. (f) Material-dependent hidden-pathway spectral response for $Ti_3C_2O_2$, $Ti_3C_2F_2$, $Ti_3C_2(OH)_2$, and $Mo_2TiC_2O_2$.

The hidden inter-orbital dynamics in Figure 11 are resolved simultaneously in time, transition-energy space, and pathway amplitude. The six leading transfer channels occupy distinct regions of the time–$\Delta E_{ij}$ landscape and exhibit markedly different temporal flux profiles. The $d_{xz} \rightarrow d_{x^2-y^2}$ channel develops pronounced flux maxima at intermediate times and across a broad range of energy differences, whereas the $d_{xy} \rightarrow d_{xz}$ pathway displays strong oscillatory transport concentrated predominantly at lower $\Delta E_{ij}$. The remaining channels contribute through temporally localized flux structures at characteristic energy ranges. Hidden orbital energy transport is therefore distributed across a sequence of pathway-specific time–energy windows rather than through a single dominant energy scale. Coherence associated with the dominant pathways is resolved at their characteristic energy separations. The coherence amplitudes show distinct onset times, temporal extents, and decay profiles across the six pathways. Low-energy channels are concentrated at earlier stages of the displayed evolution, whereas several intermediate- and high-energy pathways retain substantial coherence over later intervals. Pathway-resolved coherence is thus structured in both time and energy, with each transfer channel following a distinct temporal profile. Pathway-resolved energy transport and coherence do not share a common temporal hierarchy. Strong flux activity and large coherence can occur over different portions of the time–energy domain, and the characteristic energy separation alone does not determine the temporal persistence of either quantity. The combined representation resolves the hidden dynamics as a pathway-dependent structure in which energy transfer and inter-orbital coherence evolve along distinct yet simultaneously time-resolved channels.

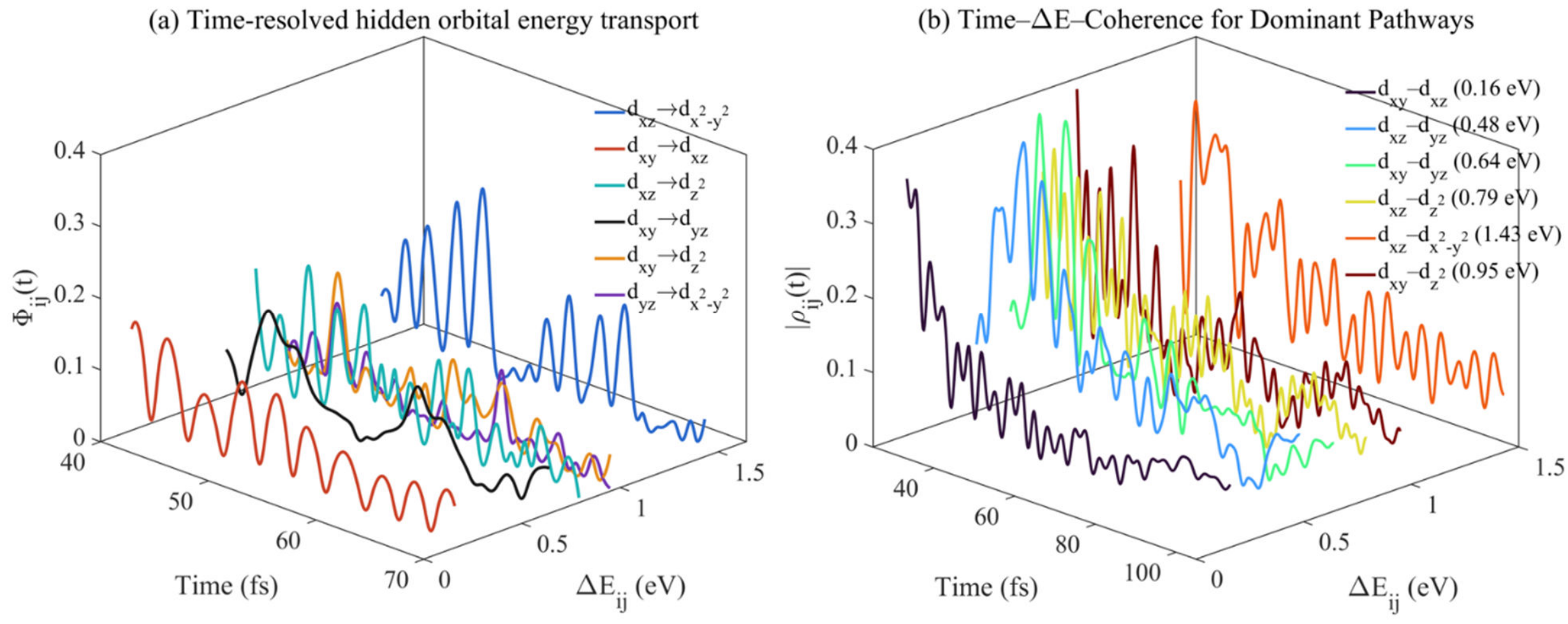

**Figure 11:** Time–energy landscapes of hidden inter-orbital transport and pathway coherence. (a) Time-resolved hidden orbital energy transport, showing the pathway-resolved flux $\Phi_{ij}$ as a function of time and inter-orbital energy difference $\Delta E_{ij}$ for the leading hidden transfer channels. (b) Time–energy distribution of the corresponding pathway coherence $|\rho_{ij}|$, resolving the characteristic energy separations and distinct temporal coherence profiles of the dominant pathways.

# IV. Conclusion

Non-thermal relaxation in MXenes is governed by a structured hierarchy of inter-orbital pathways rather than by any single population-decay or energy-dissipation coordinate. Orbital populations, cumulative transfer, instantaneous directional transfer, quantum coherence, and spectroscopic visibility furnish non-equivalent representations of the underlying dynamics and therefore expose distinct pathway hierarchies. Pathway-resolved analysis uncovers a hidden sector of the relaxation network whose dynamical weight is not necessarily mirrored by conventional spectroscopic visibility. These hidden channels display characteristic temporal, spectral, topological, and material-dependent signatures, demonstrating that microscopically important energy-transfer routes can remain only weakly expressed in directly observable signals. The resulting separation between dynamical significance and spectroscopic visibility supplies a microscopic rationale for why faint spectral features may nevertheless mediate substantial ultrafast energy redistribution. Comparison across the MXene family further establishes that pathway hierarchy is strongly material-specific. Surface termination and transition-metal substitution reshape both the orbital manifold and its coupling to the phonon bath, thereby altering the competition among population transfer, energy flow, coherence, and optical visibility. The non-thermal relaxation is more accurately viewed as a material-specific network of coupled pathways than as a universal sequence of characteristic decay times. The time–energy–coherence framework developed here recasts ultrafast relaxation as a multidimensional dynamical network that encompasses both observable and hidden channels. Beyond the MXene systems examined in this work, the approach offers a general strategy for identifying, ranking, and comparing microscopic relaxation pathways in multiband quantum materials, and it lays a foundation for pathway-selective control of ultrafast energy redistribution through deliberate tuning of composition, coherence, and network connectivity.

## Acknowledgements

Some of the authors gratefully acknowledge the financial support provided by the Iran National Science Foundation (INSF) under Grant No. 4030055. Another author gratefully acknowledges the financial support provided by the National Natural Science Foundation of China (NSFC) under Grant No. W2412113.

## Author Contribution

Ali Asghar Molavi Choobini conceived the study, developed the theoretical framework, performed the analytical and numerical work, analyzed the results, and wrote the manuscript. Abbas Chimeh contributed to discussions of the theoretical framework and physical interpretation of the results, and critically reviewed the manuscript. Jinhui Zhong contributed to discussions of the material properties and physical interpretation of the results, and critically reviewed the manuscript. All authors discussed the results and approved the final version of the manuscript.

## Funding

The authors have not disclosed any funding.

## Data Declarations

The author declares no conflicts of interest.

## Data Availability

The data and computational codes supporting the findings of this study are available from the corresponding author upon reasonable request.